\documentclass{elsarticle}
\usepackage{hyperref}
\usepackage[utf8]{inputenc}
\usepackage{wrapfig}
\usepackage{amsmath}
\usepackage{amsthm}
\usepackage{amssymb}
\usepackage[tmargin=1in,bmargin=1in]{geometry}
\usepackage{caption}
\usepackage{graphicx}
\usepackage{latexsym}
\usepackage{pdfsync}
\usepackage[boxed]{algorithm}
\usepackage{algorithm}
\usepackage{algpseudocode}
\usepackage{multirow}
\usepackage{rotating}
\usepackage{color}
\usepackage{url}
\usepackage{subfigure}
\usepackage{bigstrut}
\usepackage{bigints}
\usepackage[normalem]{ulem}

\hypersetup{
    colorlinks=true,              
    linkcolor=blue,               
    citecolor=blue,               
    filecolor=black,              
    urlcolor=red,               
    bookmarks=false,
    pdffitwindow=true,
    pdfpagelayout=SinglePage
}

\definecolor{mypink}{cmyk}{0, 0.7808, 0.4429, 0.1412}

\definecolor{mybrown}{cmyk}{0, 0.60, 0.95, 0.63}
\definecolor{darkgreen}{cmyk}{0.98, 0, 0.36, 0.22}

\newcommand{\etal}{\textit{et al.\ }}
\newcommand{\ie}{\textit{i}.\textit{e}., }
\newcommand{\eg}{\textit{e}.\textit{g}. }

\newcommand{\vect}[1]{\boldsymbol{#1}}

\newcommand{\be}{\begin{equation}}
\newcommand{\ee}{\end{equation}}
\newcommand{\ben}{\begin{equation*}}
\newcommand{\een}{\end{equation*}}
\newcommand{\bea}{\begin{eqnarray}}
\newcommand{\eea}{\end{eqnarray}}
\newcommand{\bean}{\begin{eqnarray*}}
\newcommand{\eean}{\end{eqnarray*}}

\newcommand{\pforest}{\texttt{p4est }}
\newcommand{\voropp}{\texttt{Voro++ }}

\begin{document}

\title{A parallel Voronoi-based approach for mesoscale simulations of cell aggregate electropermeabilization}

\cortext[cor]{Corresponding author: pouria@ucsb.edu}

\address[MECHE]{Department of Mechanical Engineering, University of California, Santa Barbara, CA 93106-5070}
\address[CS]{Department of Computer Science, University of California, Santa Barbara, CA 93106-5110}
\address[BIOI]{Team MONC, INRIA Bordeaux-Sud-Ouest, Institut de Math\'ematiques de Bordeaux, CNRS UMR 5251 \&
Universit\'e de Bordeaux, 351 cours de la Lib\'eration, 33405 Talence Cedex, France.
}
\author[MECHE]{Pouria Mistani}
\author[MECHE]{Arthur Guittet}
\author[BIOI]{Clair Poignard}
\author[MECHE,CS]{Frederic Gibou}

\begin{abstract}
We introduce a numerical framework that enables unprecedented direct numerical studies of the electropermeabilization effects of a cell aggregate at the meso-scale. Our simulations qualitatively replicate the shadowing effect observed in experiments and reproduce the time evolution of the impedance of the cell sample in agreement with the trends observed in experiments. This approach sets the scene for performing homogenization studies for understanding the effect of tissue environment on the efficiency of electropermeabilization.  We employ a forest of Octree grids along with a Voronoi mesh in a parallel environment that exhibits excellent scalability. We exploit the electric interactions between the cells through a nonlinear phenomenological model that is generalized to account for the permeability of the cell membranes. We use the Voronoi Interface Method (VIM) to accurately capture the sharp jump in the electric potential on the cell boundaries. The case study simulation covers a volume of $(1\ mm)^3$ with more than $27,000$ well-resolved cells with a heterogeneous mix of morphologies that are randomly distributed throughout a spheroid region.
\end{abstract}


\maketitle

\section{Introduction}
\label{sec::introduction}
Electropermeabilization (also called electroporation) is a significant increase in the electrical conductivity and permeability of the cells' membrane that occur when pulses of large amplitude (a few hundred volts per centimeter) are applied. The physical basis of this phenomenon lies in the fact that, since membranes are mainly composed of phospholipids and proteins, they behave like a capacitor in parallel with a resistor. The applied electric field is then dramatically enhanced in the vicinity of the membrane, leading to a jump of the electric potential. This locally varying transmembrane potential difference (TMP) can prevail over the cell membrane barrier in regions where this difference surpasses the electroporation threshold.

This phenomenon has attracted increasing attention due to its capacity to facilitate targeted drug delivery of non-permeant cytotoxic molecules such as bleomycin or cisplatin \cite{belehradek1994electropermeabilization}. DNA vaccination and gene therapy are other promising applications for electropermeabilization, which enables non-viral gene transfection~\cite{mcmahon2001}.

However, despite extensive scrutiny of this phenomenon, no substantial evidence of the elementary mechanism of electropermealization has been obtained. Nevertheless, the most accepted theory speculates the creation of pores in the membrane as a consequence of a large transmembrane voltage. However these pores have not yet been observed. One important reason behind this inability is that, in the absence of cell imaging techniques in the nanometer scale, almost all experiments that have studied the electroporation effect have used tissue scale samples to infer the underlying molecular level processes.

Such inferences have led to the advent of different theoretical models, with membrane pore density approaches being among the most popular mechanisms. Developments in this avenue have been carried out in the work of Debruin and Krassowska \cite{DEBRUIN} and have been augmented in \cite{KRASSOWSKA2007404} and \cite{LI201110} to incorporate the spatio-temporal evolution of the speculated pore radii. Other attempts have been made to model the tissue scale behavior of electropermeabilization \cite{Langus2016}.

Recently, Leguebe \etal \cite{LEGUEBE201483} have proposed a phenomenological approach to model this effect at single-cell scale in terms of a nonlinear partial differential equation. Their description determines the local behavior of each cell membrane under the influence of its surrounding electric potential in a continuous manner. Remarkably, this representation qualifies for a multi-scale characterization of electropermeabilization. However, we note that in practice these models embody calibrations of free parameters that are tuned by experimenting on populations of cells and extending these measurements to single-cell scale, oversighting the multi-scale nature of electropermeabilization in the experiments. Such approximations are inevitable in the absence of numerical tools to adjust these models in accordance with experiments. However, recent attempts have been made in the work of Voyer \etal \cite{VOYER201898} to theoretically extend this model to tissue scale.

We emphasize the predictability of any such model at the cell aggregate regime to corroborate these results. However, such comparisons with available experimental results were prohibitive in the case of electropermeabilization, partially due to the enormous computational costs of such ventures as well as the complexity of the molecular events involved in membrane electropermeabilization. To facilitate the accurate modeling of molecular processes that regulate electropermeabilization, there has been emerging incentive to overcome the hindering computational difficulties. 

In the wake of the aforementioned arguments, the advent of ``direct'' tissue scale simulations seems necessary. Such simulations not only commission better understanding of the involved molecular processes, but also will aid developing semi-analytic models of the overall permeabilization of the tissue under different circumstances. Such endeavors require a complete characterization of the relevant physical parameters from cell scale physics to tissue scale configurations.

Quite recently, significant progress has been made in this venue by Guittet \etal \cite{guittet2017voronoi}. They have proposed a novel Voronoi Interface Method (VIM) to capture the irregular cell interface and accurately impose the sharp TMP jump. The VIM utilizes a Voronoi mesh to capture the irregular interface before applying the dimension-by-dimension Ghost Fluid Method \cite{FEDKIW1999457,Kang2000,liu2003convergence}. This is aimed to direct the fluxes normal to the interface where there is a discontinuity. This reframing the mesh around the interface guarantees the convergence of the solution's gradients. Also, only the right hand side is affected by the TMP jump which simplifies the computational treatment. 

We also note that an alternative framework would be using adaptive Chimera grids as proposed by English \etal \cite{english2013adaptive}. In their proposed method, English \etal used multiple Cartesian grids in different regions of the domain that are coupled on their boundaries by generating a Voronoi mesh. In the case of electroporation, one could also use finer Cartesian grids near the cell membrane that are coupled on the cell boundary with a Voronoi extension. 

Guittet \etal \cite{guittet2017voronoi} have derived a finite volume discretization for this phenomenon and implemented it in a serial framework. Their numerical results are in agreement with experimental expectations. However, the computational costs of solving the involved discretization prohibited the consideration of tissue scale simulations.

Here, we build on the method proposed by Guittet \etal \cite{guittet2017voronoi} and generalize their approach to a parallel environment. This parallelization empowers simulations of the single-cell model of Legu\`ebe and Poignard \etal \cite{LEGUEBE201483} at the tissue scale, hence providing a framework to validate or improve the understanding of cell electroporation.

The structure of this paper is as follows. We introduce the mathematical model for our simulations in section \ref{sec::cell_membrane_model} and the computational strategy that we develop in section \ref{sec::parallelAdaptiveStrategy}. Then we present performance of our implementation as well as some preliminary demonstrations of the numerical results in sections \ref{sec::NumericalResults}. In section \ref{sec::emergent} we illustrate the emergence of macro-level properties in the cell aggregate. We conclude with a summary of our main results in section \ref{sec::Conclusion}.


\section{Cell membrane model}
\label{sec::cell_membrane_model}
\subsection{Geometric representation}
\label{subsec::Geometric_rep}
\begin{figure}[H]
\begin{center}
\subfigure{\includegraphics[height=.2\textwidth]{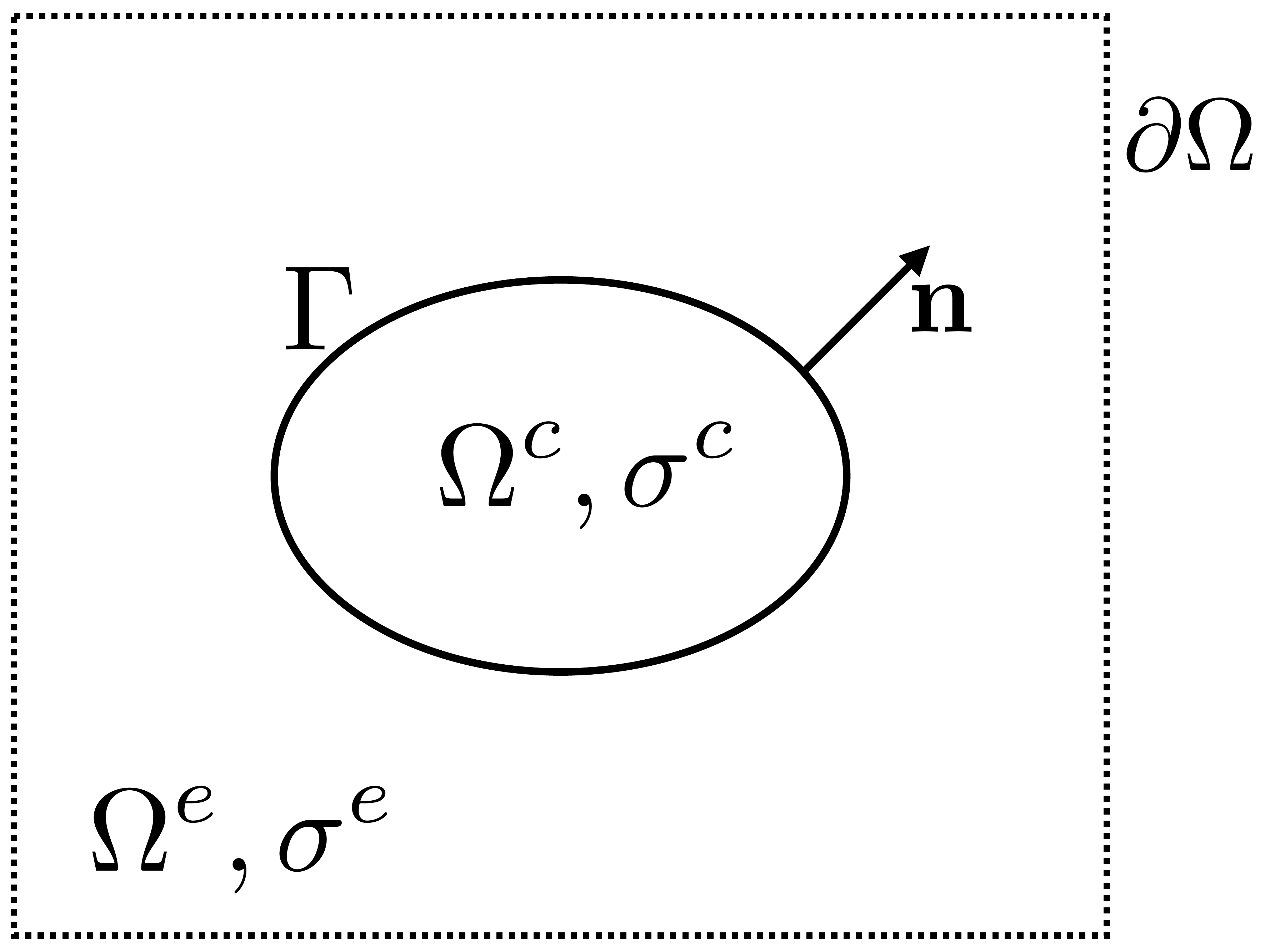} \label{subfig::geometry}} \quad \quad
\end{center}
\caption{\it Illustration of a single cell immersed in the extracellular matrix. The conductivity of the materials is denoted by $\sigma$.} \label{fig::geometry}
\end{figure}

The cell cytoplasm $\Omega^c$ and the extracellular matrix $\Omega^e$ are separated by a thin and resistive membrane denoted by $\Gamma$. The outward normal to $\Omega^c$ is denoted by $\mathbf{n}$. Figure \ref{fig::geometry} illustrates the geometry in the case where a single cell is considered. The entire domain is denoted by $\Omega = \Omega^e \cup \Gamma \cup \Omega^c$. We denote the conductivities of the materials by $\sigma^c$ and $\sigma^e$ for the cell and the extracellular matrix respectively.

\subsection{Electrical model}
\label{subsec::Elect_model}
For simulating the electropermeabilization process, we solve the following boundary value problem defined in equations ~\eqref{eq::Laplace}--\eqref{eq::IC}. The electric potential field $u$ in the computational domain is governed by the Laplace equation:
\begin{subequations}
\begin{align} 
&\Delta u = 0, \  \vect{x} \in (\Omega_c \cup \Omega_e), \label{eq::Laplace}
\intertext{with the appropriate boundary conditions:}
&\left[\sigma \partial_{\vect n} u\right]_\Gamma = 0,\ \vect{x} \in \Gamma, \label{eq::bc1} \\
&C_m \partial_t \left[u\right]_\Gamma + S(t,\left[u\right])\left[u\right] = \sigma \partial_{\vect n} u\vert_\Gamma, \ \vect{x} \in \Gamma,\label{eq::bc2} \\
&u(t,\vect{x}) = g(t,\vect{x}),\ \vect{x} \in \partial\Omega, \label{eq::bc3} 
\intertext{and the homogeneous initial condition:}
&u(0,\vect{x}) = 0, \ \vect{x} \in \Omega, \label{eq::IC}
\end{align}
\end{subequations}
where we used the $\left[ \boldsymbol{\cdot}\right]$ notation for describing the jump operator across $\Gamma$.

Equation \eqref{eq::bc1} imposes the continuity of the electric flux across the membrane, \eqref{eq::bc2} captures the capacitor and resistor effect of the membrane and \eqref{eq::bc3} is the external voltage applied on the boundaries of the domain. In these equations, $C_m$ and $S$ are the capacitance and conductance of the membrane material respectively. The source term corresponding to the applied voltage is denoted by $g(t,\vect{x})$. The effect of the electroporation current is modeled by the $S(t,\left[u\right])\left[u\right]$ term in equation \eqref{eq::bc2}. We adopt a nonlinear description of the conducting membrane \cite{LEGUEBE201483} in the next subsection.
 
\subsection{Membrane electropermeabilization model}
\label{subsec::Electroperm_model}
The long-term permeabilization of the membrane is modeled by formulating the surface membrane conductivity. Leguèbe, Poignard \etal \cite{LEGUEBE201483} modeled the surface conductivity of the membrane as follows:
\begin{equation}
S_m(t,s) = S_0 +S_{ep}(t,s) = S_0 +X_1(t,s)\times S_1+X_2(t,s)\times S_2,\ \ \  \forall t>0, s\in\Gamma \label{eq::conductance}
\end{equation}

In this equation  $S_0$, $S_1$ and $S_2$ are the surface conductance of the membrane in the resting, porated and permeabilized
states, respectively. The level of poration and permeabilization of the membrane are captured in the functions $X_1$ and $X_2$. These are computed as a function of the transmembrane potential difference and are valued in the range $\left[0,1\right]$ by definition. The ordinary differential equations determining $X_1$ and $X_2$ read: 

\begin{subequations}
\begin{align}
\frac{\partial X_1(t,s)}{\partial t} = \frac{\beta_0(s) - X_1}{\tau_{ep}}, \ \ \ X_1(t,s) = 0, \label{eq::porosity} \\
 \frac{\partial X_2(t,X_1)}{\partial t} = max\bigg(\frac{\beta_1(X_1) - X_2}{\tau_{perm}}, \frac{\beta_1(X_1) - X_2}{\tau_{res}}  \bigg), \ \ \ X_2(t,s) = 0.\label{eq::perm} 
\end{align}
\end{subequations}
The parameters $\tau_{ep}$, $\tau_{perm}$ and $\tau_{res}$ are the time scales for poration, permeabilization and resealing, respectively. 
Furthermore, in the above equations $\beta_0$ and $\beta_1$ are regularized step-functions defined by:
\begin{subequations}
\begin{align}
\beta_0(s) = e^{-\frac{V_{ep}^2}{s^2}}, \ \ \ \forall s \in \mathbb{R}, \label{eq::beta_0}, \\
\beta_1(X) = e^{-\frac{X_{ep}^2}{X^2}}, \ \ \ \forall X \in \mathbb{R}, \label{eq::beta_1},
\end{align}
\end{subequations}
where $V_{ep}$ and $X_{ep}$ are the membrane voltage and the poration thresholds respectively.

\section{Computational strategy} \label{sec::parallelAdaptiveStrategy}
\subsection{Level-set representation}
\label{subsec::LevelSet}

\begin{figure}[H]
\begin{center}
\subfigure{\includegraphics[height=.4\textwidth]{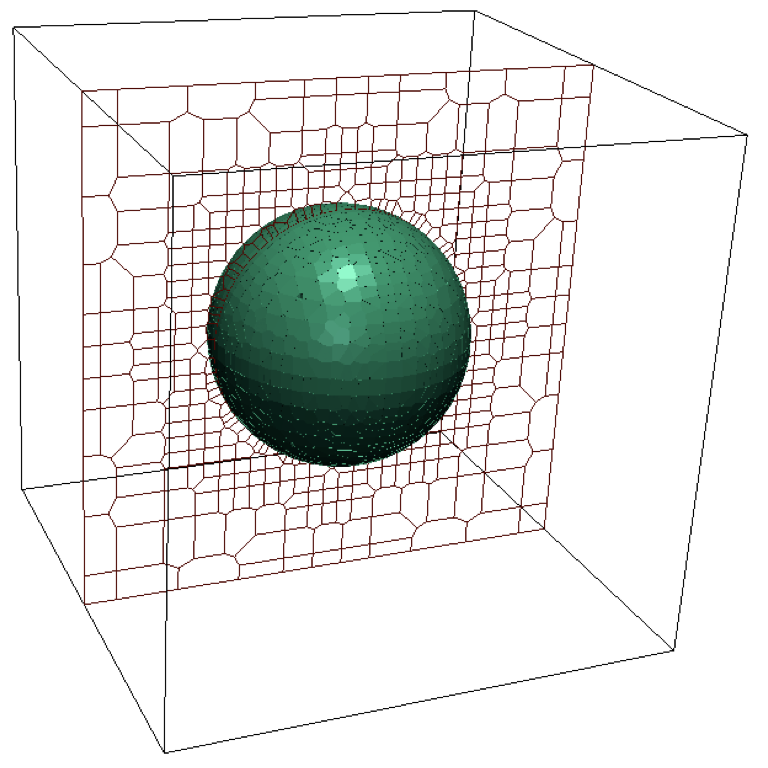} \label{subfig::cell}} \quad \quad
\subfigure{\includegraphics[height=.4\textwidth]{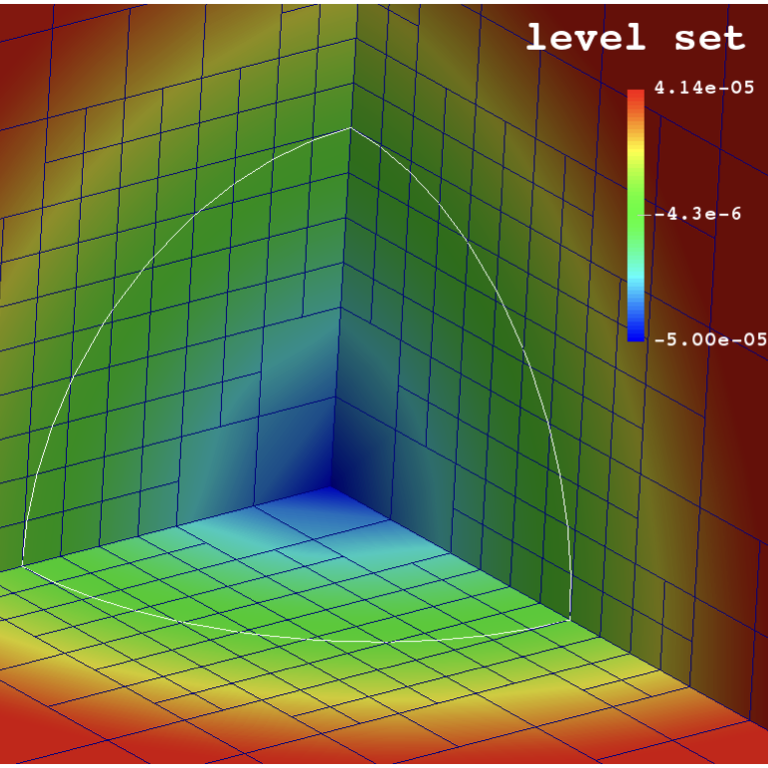} \label{subfig::phi}} \quad \quad
\end{center}
\caption{\it  (left) A Voronoi mesh is fitted to the surface of a membrane in 3D. (right)  The level-set representation of a single cell on the dual adaptive Cartesian grid at levels $(4,6)$. The membrane is resolved at the highest resolution while farther regions are at lower resolution. Also, the level-set function $\phi$ is negative inside the cell (cooler colors) and positive outside the cell (warmer colors). } 
\end{figure}

As presented by Guittet \etal \cite{guittet2017voronoi}, we describe the cells in our simulations using the level-set method as first introduced by \cite{OSHER198812} (see \cite{Gibou;Fedkiw;Osher:18:A-review-of-level-se} for a recent review) and in particular the technology on Octree Cartesian grid by Min and Gibou \cite{Min;Gibou:07:A-second-order-accur}. To this end, we construct a spatial signed-distance function $\phi$ relative to the irregular interface $\Gamma$ such that:

\begin{equation}
\phi(\vect{x}) = \begin{cases}
						\ \ d(\vect{x},\Gamma)> 0 ,\ \vect{x} \in \mathcal{O}_e\\
						\ \ d(\vect{x},\Gamma) = 0,\ \vect{x} \in \Gamma\\
						 -d(\vect{x},\Gamma) < 0 ,\ \vect{x} \in \mathcal{O}_c
						\end{cases}, \,\vect{x} \in \mathbb{R}^{3},
\end{equation}
where $d(\vect{x},\Gamma)$ is the Euclidean distance from a given point in the domain to the 0-th level-set hyperspace:
$$
 d(\vect{x},\Gamma) = \underset{\vect{y} \in \Gamma} \inf\ d(\vect{x},\vect{y}),
$$
Figures \ref{subfig::cell} and \ref{subfig::phi} give an example of such interface representation and a sample level-set function, respectively. 

\subsection{Octree data structure and refinement criterion} 
\label{subsec::OctreeDataStructure}

\begin{figure}[H]
\begin{center}
\subfigure{\includegraphics[height=.35\textwidth]{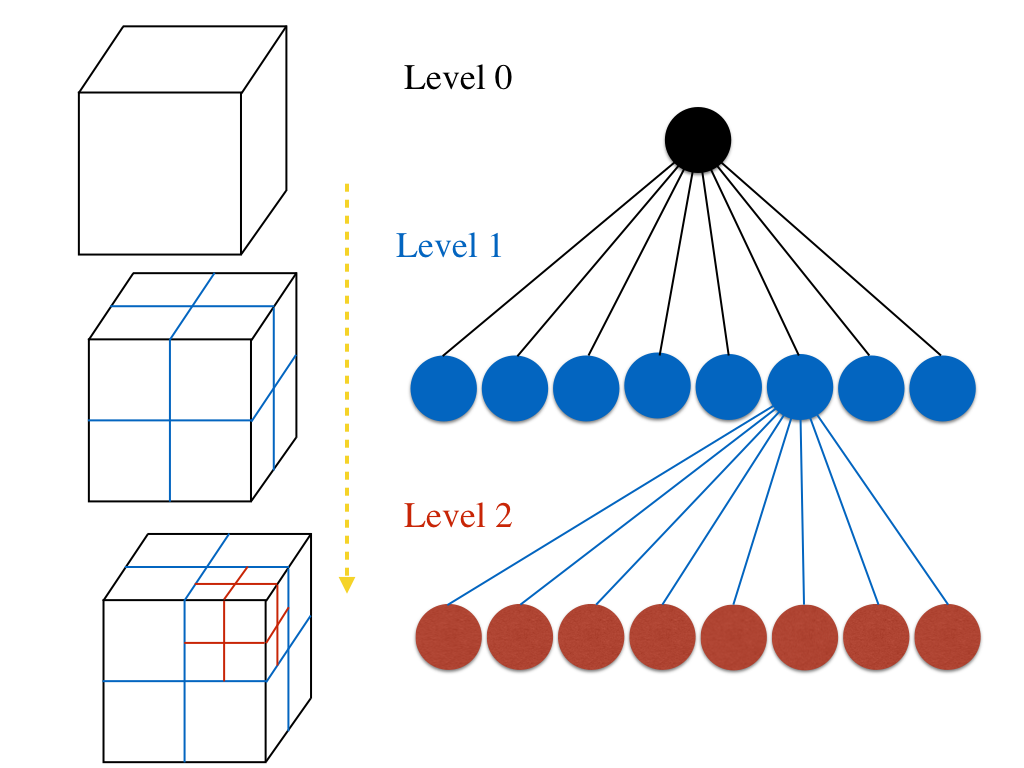} \label{subfig::Octree}} \quad \quad
\subfigure{\includegraphics[height=.35\textwidth]{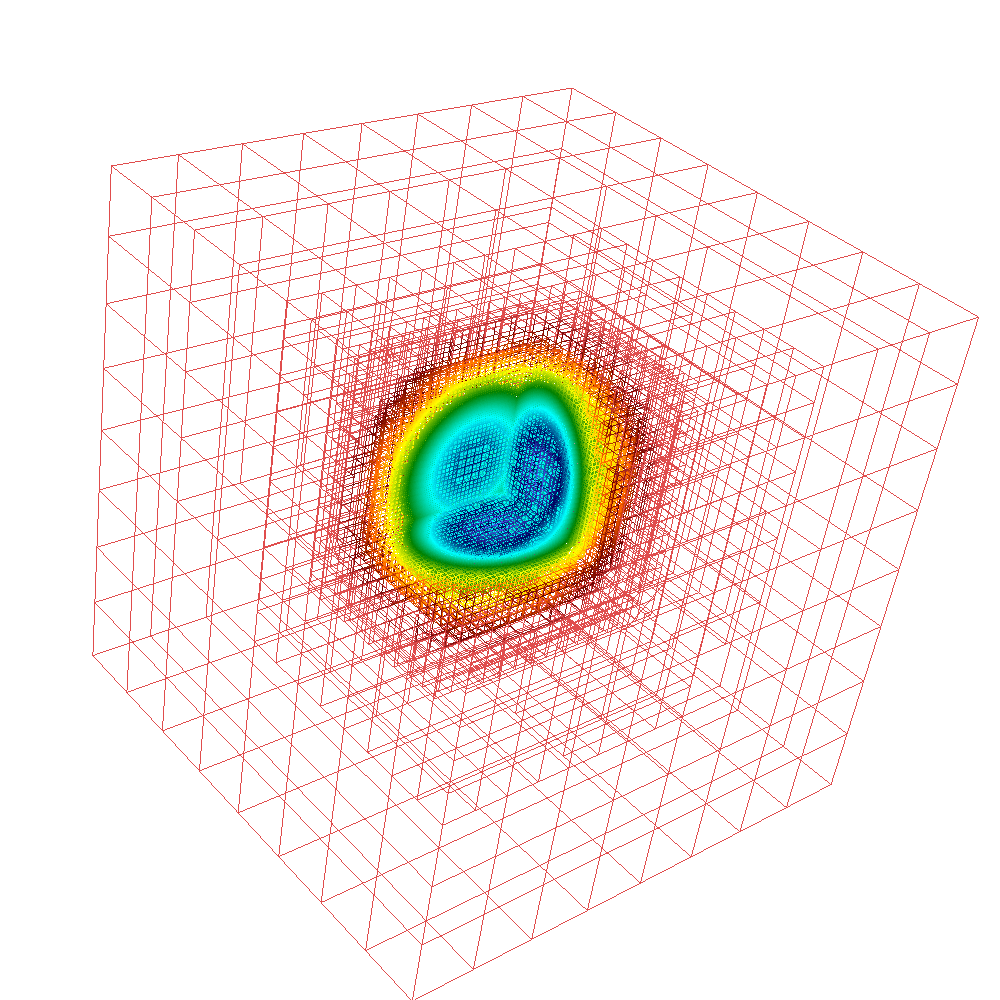} \label{subfig::full_tree}} \quad \quad
\end{center}
\caption{\it Illustration of an Octree mesh and its data structure. (left) Two levels of refinement are illustrated. (right) A portrait of 8 levels of refinement in practice. Note that each dimension is divided into at most $2^8=256$ cells.} 
\end{figure}

Simulating a large number of biological cells in three spatial dimensions requires minimizing the total number of degree of freedom without loss of accuracy. As the physical variations in the solution occur close to the membrane, one needs more nodes to capture the physics at the vicinity of the biological cells compared to farther regions. We utilize the adaptive Cartesian grid based on Quad-/Oc-trees \cite{Finkel1974,MEAGHER1982129}. A ``Quad-/Oc-tree'' is a recursive tree data structure where each node is either a leaf node or a parent to 4/8 children nodes. The Octree is constructed by setting the root of the Octree to the entire computational domain. Then higher resolutions are achieved by recursively dividing each cell into 8 subcells (or 4 subcells in the case of Quadtrees). We use the following refinement criteria introduced by \cite{strain1999tree} and extended by \cite{min2004local} to orchestrate this partitioning of space:

\textbf{Refinement/coarsening criterion:} Split a cell ($\mathcal{C}$) if the following inequality applies (otherwise merge it to its parent cell):
\begin{equation}
\min_{v\in \textrm{vertices}(\mathcal{C})}|\phi(v)| \le \textrm{Lip}(\phi)\cdot \textrm{diag-size}(\mathcal{C}),
\label{eq::refinement}
\end{equation}
where we choose a Lipschitz constant of Lip$(\phi)\approx 1.2$ for the level-set $\phi$. Furthermore, diag-size($\mathcal{C}$) stands for the length of the diagonal of $\mathcal{C}$ and $v$ refers to its vertices. Intuitively, the use of the signed-distance function in equation \eqref{eq::refinement} translates into a refinement based on distance from the interface. This process is depicted in figure \ref{subfig::Octree}. An Octree is then characterized by its minimum/maximum levels of refinement. Figure \ref{subfig::full_tree} illustrates an example of a levels (3,8) tree meaning the minimum and maximum number of cells in each dimension are $2^3=8$ and $2^8=256$ respectively.
Note that if larger macromesh is used these numbers will be multiplied by the macromesh number; \eg if one sets $n_x=2$ for levels $(3,8)$ then the number of cells in $x$-direction will be twice as before, \ie bound between $16$ and $512$ instead. This is the case in all of the simulations in this work.

\subsection{Parallel framework}
\label{subsec:parallel-frame}
\begin{figure}[hbtp]
\begin{center}
\fbox{\includegraphics[height = .11\textheight]{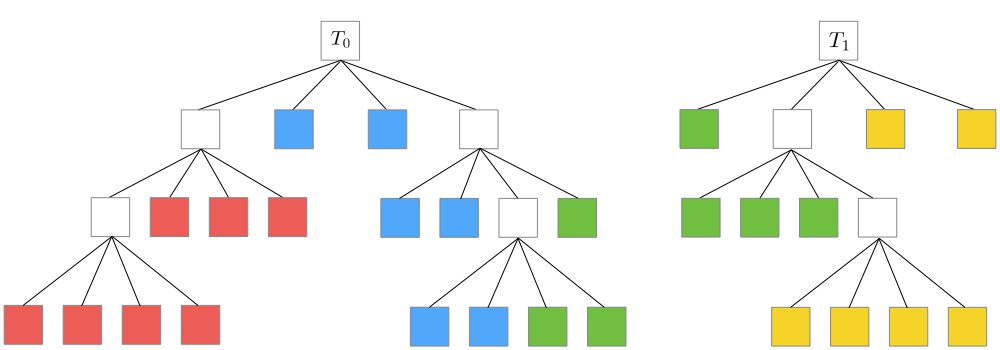}} \quad \fbox{\includegraphics[height = .11\textheight]{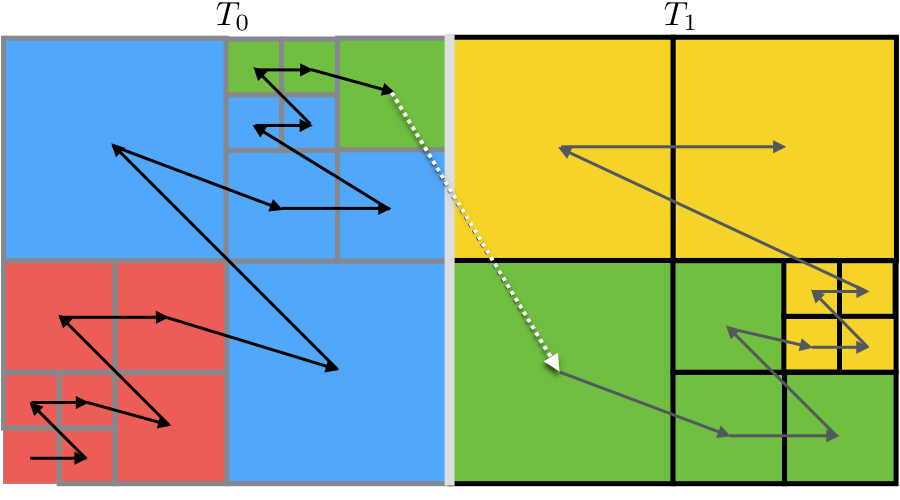}} \quad \fbox{\includegraphics[height = .11\textheight]{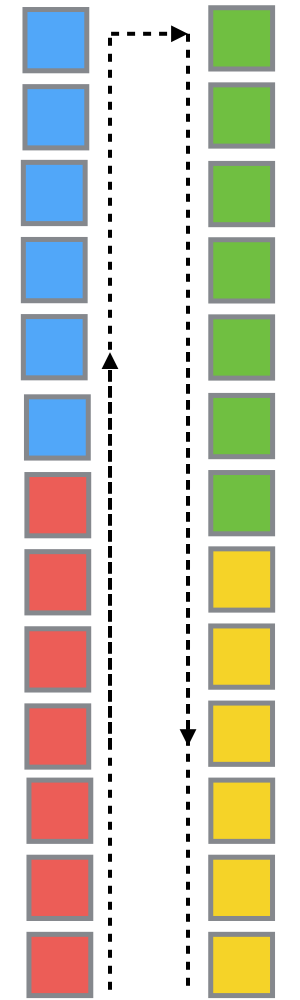}}
\caption{ A ``forest'' composed of two Quadtrees $T_0$ and $T_1$ (left) partitioning the whole geometric domain following a $Z$-ordering of all the octants in the domain (center). The partitioning is performed such that each process receives equal ($\pm1$) number of contiguous octants traversing the leaves from left to right (right). Here there are four different processes depicted by four different colors.}
\label{fig::p4est_partitioning}
\end{center}
\end{figure}

We utilize the parallelism scheme introduced by Mirzadeh \etal \cite{Mirzadeh2016345}. This scheme is built upon the $\pforest$software library \cite{burstedde2011p4est}. $\pforest$ is a suite of scalable algorithms for parallel adaptive mesh refinement/coarsening (AMR) and partitioning of the computational domain to a forest of connected Quad-/Oc-trees. The partitioning strategy used in \pforest is illustrated in figure \ref{fig::p4est_partitioning}. This process is \cite{burstedde2011p4est}:
\begin{itemize}
\item A uniform macromesh is created;
\item A forest of Octrees is recursively constructed using all processes; 
\item The produced tree is partitioned among all processes using a $Z$-ordering; \ie a contiguous traversal of all the leaves covering all the octrees.
\end{itemize}

The $Z$-ordering is then stored in a one dimensional array and is equally divided between the processes. This contiguous partitioning optimizes the communication overhead compared to the computation costs when solving equations in parallel. To perform the discretizations derived for this problem, we need to construct the local Octrees from the one dimensional array of leaves. To this end, following the method suggested by Mirzadeh \etal \cite{Mirzadeh2016345}, we construct a local tree on each process such that the levels of its leaves matches that of the leaves produced by the $\pforest$ refinement. This is because $\pforest$ does not provide the vertical structure, and we need to be able to find a cell containing a point quickly, in $\mathcal{O}(\log(N))$. Each process stores only its local grid plus a surrounding layer of points from other processes, \ie a ghost layer.

\subsection{Quasi-random cell distribution}
\label{subsec:quasirandom}
To computationally capture the effects of a large aggregate of cells under the influence of an external electric stimulant, first we need to efficiently mimic the randomness in the distribution of the cells while simultaneously constraining the minimum distance among the cells. In fact, for the purposes of this work we need to simulate tens to hundreds of thousands of cells in a relatively small computational domain if we are to observe the relevant aspects of electropermeabilization at the tissue scale.

\begin{figure}[H]
\begin{center}
\subfigure{\includegraphics[height=.45\textwidth]{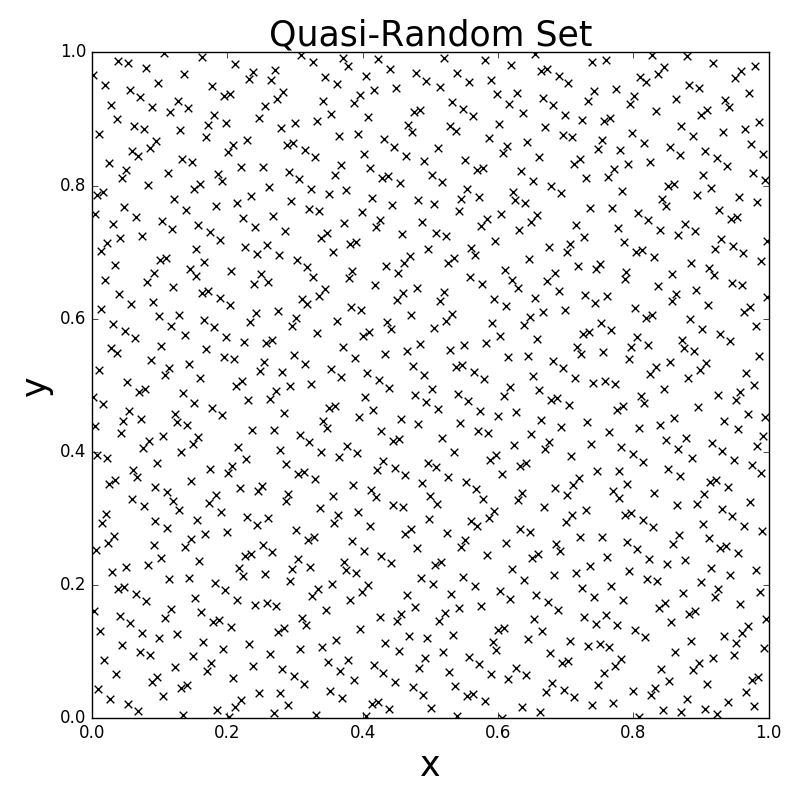} \label{subfig::halton}} \quad \quad
\subfigure{\includegraphics[height=.45\textwidth]{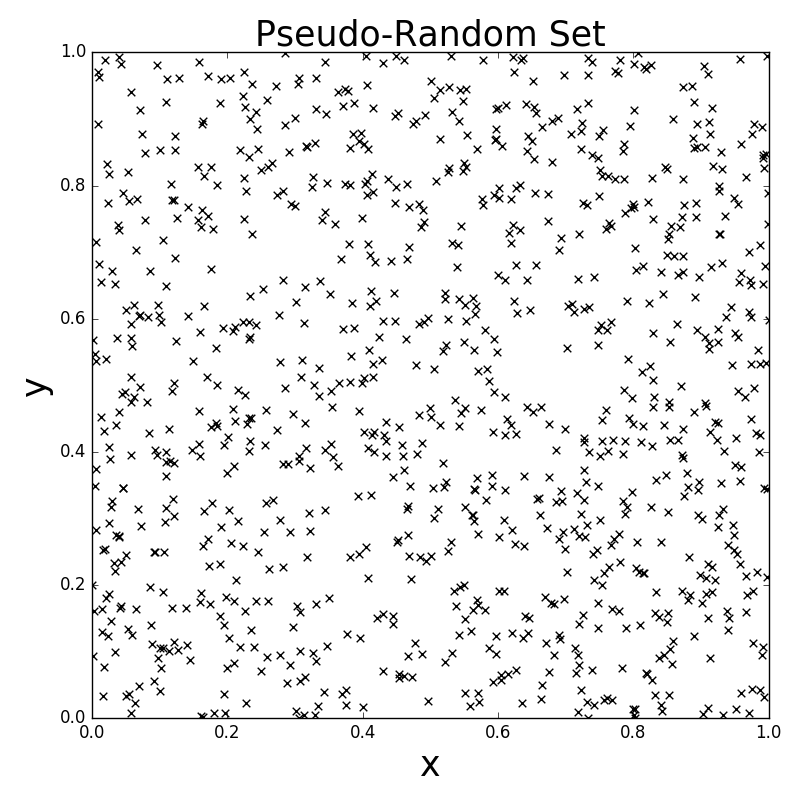} \label{subfig::uniform}} \quad \quad
\end{center}

\caption{\it (left) Quasi-random number distribution versus (right) pseudo-random number distribution. The quasi-random sequence immediately exhibits a much more uniform distribution of points.} \label{fig::random}
\end{figure}

To this end, we distribute the cells using the quasi-random numbers generated by the Halton Quasi Monte Carlo (HQMC) sequence \cite{levy2002introduction,Halton1960,BRAATEN1979249,ArtScientificComputing}. Quasi-random sequences are more uniformly distributed than the well-known pseudo-random sequences as illustrated in figure \ref{fig::random}. As seen in this figure, while uniform pseudo-random  numbers suffer from local clustering and voids, the HQMC sequence spans the space more uniformly. Mathematically, the uniformity of a sequence is measured by its ``discrepancy'' which is measured by comparing the number of points in a given region of space with the number of points expected from an ideal uniform distribution \cite{levy2002introduction}. The quasi-random sequences are also called \textit{low} discrepancy sequences as they exhibit a more uniform spatial coverage. Remarkably, the low discrepancy characteristic is inherently built in the HQMC algorithm, as opposed to a pseudo-random number generator that would require further processing.

In our approach, we locate each cell at the next element in a three dimensional HQMC sequence while skipping the elements that violate the minimum distance criterion to the previously located cells. In contrast to a pseudo-random based technique, such rejections are very rare due to the intrinsic low discrepancy of the HQMC sequence, and hence the efficiency of our technique. As the number of cells increases in our simulations, it becomes computationally prohibitive to generate such a non-overlapping pseudo-random distribution of cells at high densities. Our experiments with HQMC demonstrate that a moderately dense non-overlapping cluster of cells can be generated at least hundreds of times faster than a pseudo-random number based technique. Notably, initializing higher cluster volume fractions (a volume fraction of $\rm n = \dfrac{volume\ of\ the\ cells}{volume\ of\ the\ spheroid}\approx\mathcal{O}(10^{-1})$ ) seems completely impossible using pseudo-random number generators.

\subsection{Discretization of the equations - the Voronoi Interface Method}
\label{subsec:discretize}

\begin{figure}[H]
\begin{center}
\includegraphics[width=\textwidth]{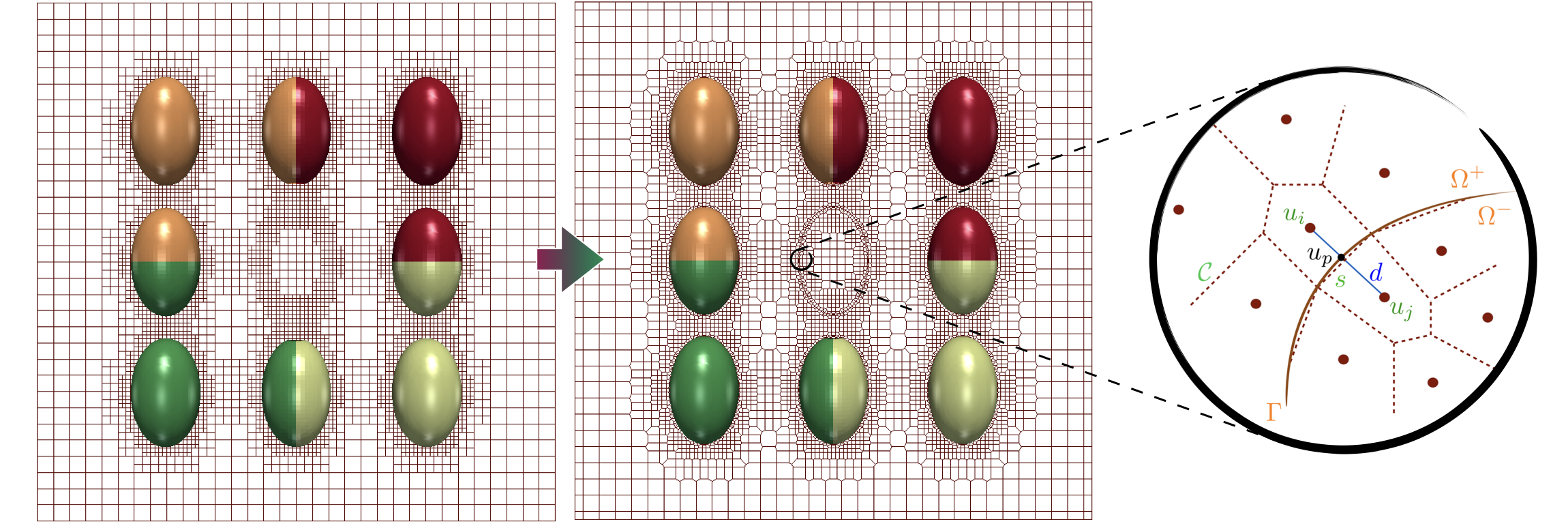} \label{subfig::voro}
\end{center}
\caption{\it (left to middle) An Octree is converted into an adaptive Voronoi mesh such that Voronoi faces are fitted to the interface. In our framework the computational domain is partitioned among different processors as demonstrated by different cell colors. (right) In our discretization, $u_p$ corresponds to the normal projection of nodes $i$ and $j$ on the interface ($\Gamma$). This point is equidistant to nodes $i$ and $j$. $s$ is the common length (or area in 3D) of the interface between cells $i$ and $j$. $d$ is the distance between $i$ and $j$.  } \label{fig::voronoi}
\end{figure}

The main difficulty in solving the equations of section \ref{subsec::Elect_model} is related to the non-trivial boundary conditions and discontinuities across the cells' surface. Guittet \etal \cite{GuittetVoronoi} introduced the Voronoi Interface Method (VIM) to solve elliptic problems with  discontinuities on irregular interfaces. Their proposed method exhibits second order accuracy by solving the problem on a Voronoi mesh instead of the given Cartesian grid. Also, Guittet \etal \cite{guittet2017voronoi} extended the VIM to the case of the electropermeabilization problem including the aforementioned non-trivial boundary condition in the discretization. In this work, we implement their modified approach in parallel. In this section we briefly highlight this technique.

The solver presented by Guittet \etal \cite{GuittetVoronoi} is based on building a Voronoi mesh using the freely available library \voropp \cite{voro}. The Poisson equation is then solved on a Voronoi mesh that coincides with the irregular interface. This introduces additional degrees of freedom close to the interface and on either side that are equidistant to the interface by design.  Briefly, the procedure for converting an initial adaptive Cartesian mesh to a conforming Voronoi mesh starts by adopting the Cartesian nodes as cell centers, \textit{i.e.} known as Voronoi seeds, for a Voronoi mesh covering the computational domain. Next, if a Voronoi cell crosses the interface we replace the corresponding degree of freedom with a pair of equidistant points on either sides of the interface. This procedure provides a conforming Voronoi tessellation of the domain such that interfaces are tiled with collections of faces from adjacent Voronoi cells. For more details on generating the Voronoi mesh we refer the interested reader to \cite{GuittetVoronoi}. Here, we present the numerical scheme of Guittet \etal \cite{guittet2017voronoi} for completeness using the nomenclature given in Figure \ref{fig::voronoi}.

First we discretize the boundary condition \eqref{eq::bc2} using a standard Backward Euler scheme:
\begin{equation}
C_m\frac{\left[u\right]^{n+1} - \left[u\right]^{n}}{\Delta t} + S^n\left[u\right]^{n+1} = (\sigma\partial_n u^{n+1})_\Gamma,
\end{equation} 
which can be rearranged to get the membrane voltage jump:
\begin{equation}
\left[u\right]^{n+1} = \frac{C_m\left[u\right]^n+\Delta t(\sigma\partial_n u^{n+1})_\Gamma}{C_m+\Delta t S^n}, \label{eq::jump}
\end{equation}
In the second step, we discretize the continuity in the electric flux boundary condition \eqref{eq::bc1}:
\begin{equation}
\sigma^e\frac{u^e_p - u^e_i}{d/2}=\sigma^c\frac{u^c_j - u^c_p}{d/2},
\end{equation}
Replacing $u^c_p$ by its definition $u^e_p-\left[u\right]^{n+1}$ in the above expression, coupling it with equation \eqref{eq::jump} and rearranging the terms, the final expression of $u^e_p$ reads:
\begin{equation}
u^e_p = \bigg(\sigma^e u^e_i + \sigma^c u^c_j + \frac{\sigma^c C_m \left[u\right]^n}{C_m+\Delta t S^n} + \frac{\sigma^c\sigma^e \Delta t }{(C_m+\Delta t S^n)d/2}u^e_i \bigg)/ \bigg(\sigma^c + \sigma^e + \frac{\sigma^c \sigma^e \Delta t}{C_m + \Delta t S^n)d/2} \bigg),
\label{eq::uep}
\end{equation}
This equation for $u^e_p$ is then included in the discretization of the Laplace equation on the Voronoi cells. Finally, we get the following expression for the potential around the interface:
\begin{equation}
\sum_{k \in \{ \partial \mathcal{C} \backslash  \Gamma\}} s_k \sigma^e \frac{u^e_k -u^e_i}{d_k} + s\hat{\sigma} \frac{u_j - u_i}{d/2} = \mathrm{sign}(\phi_i)s \hat{\sigma}\frac{C_m\left[u\right]^n}{(C_m + \Delta t S^n)d/2},
\end{equation}
where
\begin{equation}
\hat{\sigma} = \frac{\sigma^c \sigma^e}{\sigma^e + \sigma^c+\frac{\sigma^e \sigma^c \Delta t}{(C_m+ \Delta t S^n) d/2}},
\end{equation}
and ``$\rm{sign}$'' refers to the signum function.  This discretization leads to a positive definite linear system as all coefficients are positive and the jump appears only on the right-hand side of this system. We emphasize that the points far from the interface are discretized according to a standard finite volume discretization on the Voronoi grid. Integrations are performed with the geometric approach of Min and Gibou \cite{Min;Gibou:07:Geometric-Integratio}. Note that finite volume discretizations are flexible with respect to spatial variations of the Voronoi mesh topology as they only utilize values on adjacent Voronoi cell centers, as well as values of the jump on the faces midway between pairs of Voronoi cells around the interface. Despite finite difference discretizations, this aspect circumvents challenges that arise when treating the faces between coarser and finer grids.

\section{Numerical Results}
\label{sec::NumericalResults}
\subsection{Qualitative results}
First, we present numerical results illustrating the capabilities of our approach in capturing the interaction between the cell membrane and the applied electric field. Electric fields provide a feedback channel for the cell membranes to interact over long distances and leads to environmental dependence of electropermeabilization within the aggregate environment.

Second, to demonstrate this effect on a biologically relevant construct and to showcase the computational capabilities of our approach, we consider the case of a spherical aggregate of cells confined in the center of a computational box of size $1mm$  on each side. The volume fraction of cells is set to $n=0.13$ corresponding to $27,440$ well-resolved cells. The minimum distance between each pair of cells is set to  $3\times R_0$  where $R_0$ is the average radius of a cell. At present, we only intend to randomly distribute the spheroids with varying eccentricities and orientations. Therefore, this minimum threshold was adopted conservatively to avoid overlap between cells. A denser configuration would require to account for the orientation of each neighboring cell to be able to fill the free space more compactly. 

\begin{figure}[H]
\begin{center}
\subfigure{\includegraphics[height=.45\textwidth]{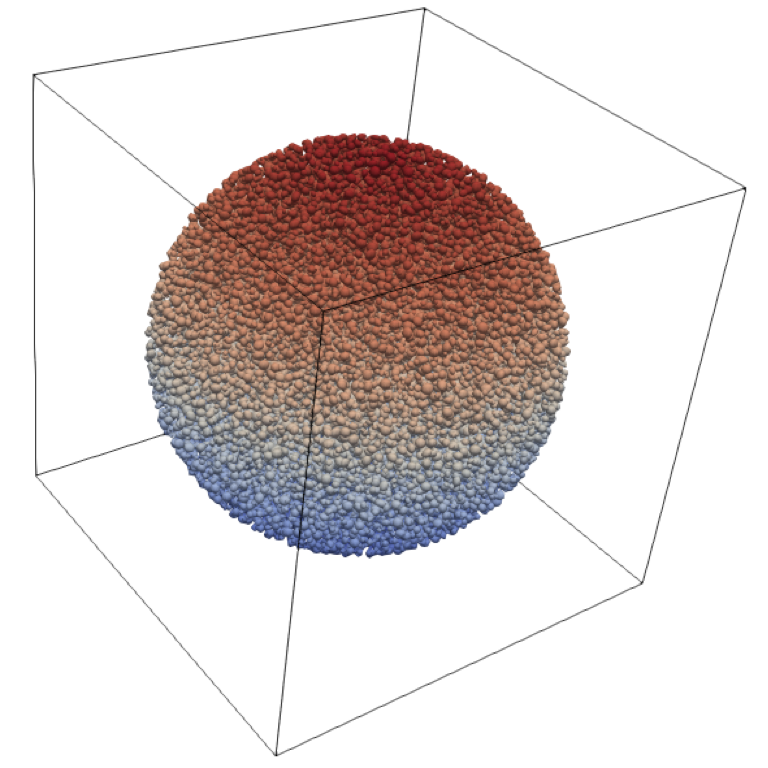} \label{subfig::clusterPhi}} \quad \quad
\subfigure{\includegraphics[height=.45\textwidth]{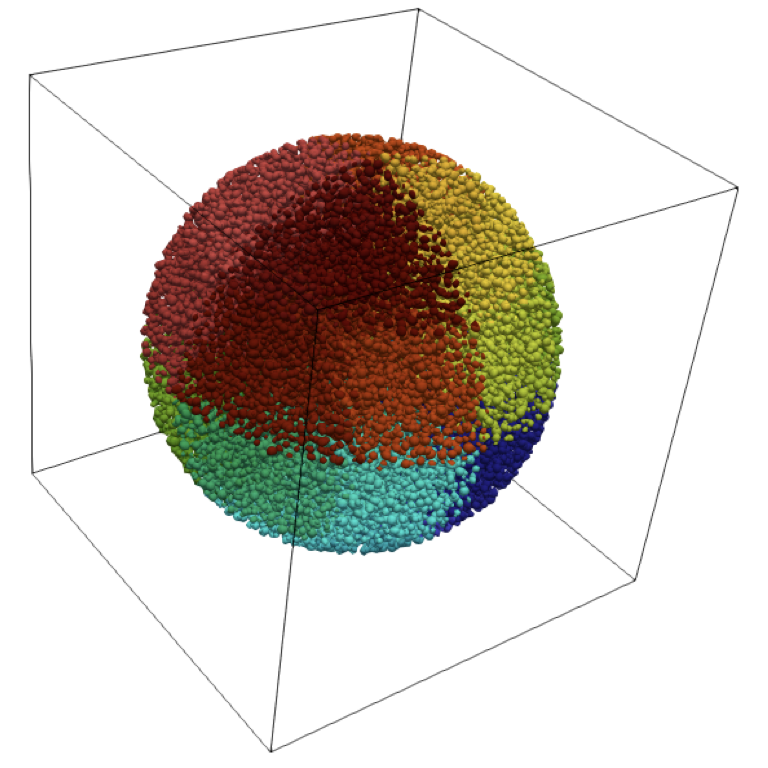} \label{subfig::clusterProcess}} \quad \quad
\end{center}
\caption{\it Illustration of a cell aggregate immersed in an external electric field. (left) colors represent the electric potential of the membranes with red being higher intensities and blue lower intensities. We note that we have set the absolute value of the bottom potential to ``0'' (ground state) while the top electrode is at our desired potential difference. (right) partitions used in this simulation. Each color represents a group of 8 processors used in this simulation.} \label{fig::clusterconfig}
\end{figure}

\begin{figure}[H]
\begin{center}
\subfigure{\includegraphics[height=0.36\textwidth]{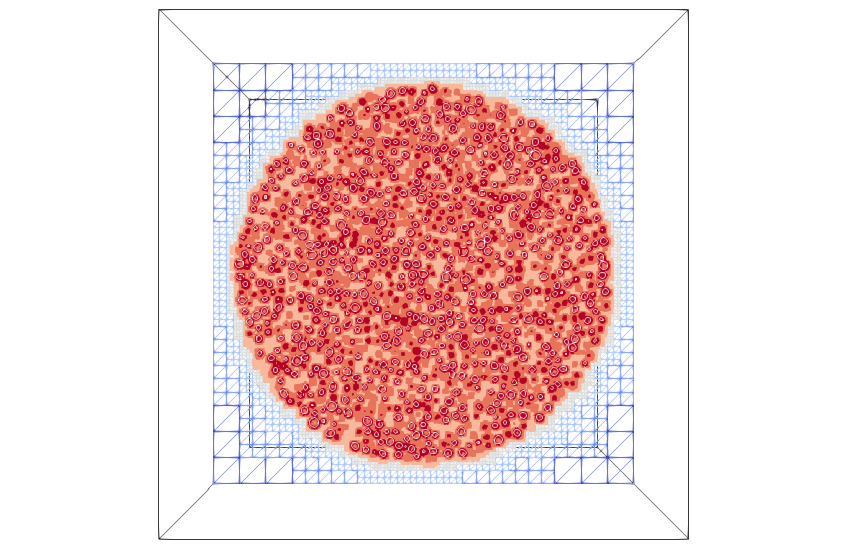} \label{subfig::zoom1}} \quad \quad
\subfigure{\includegraphics[height=0.36\textwidth]{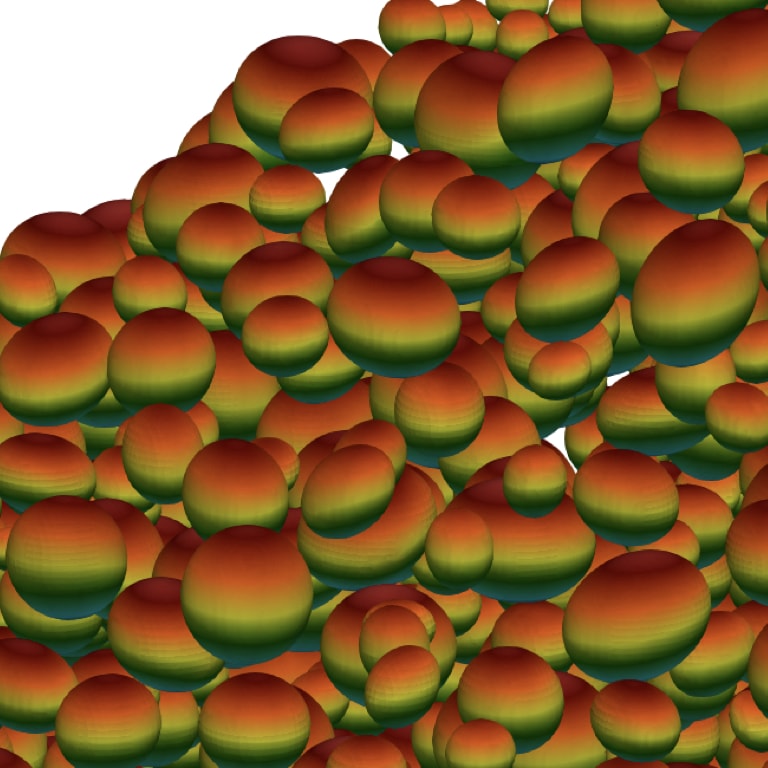} \label{subfig::zoom6}} \quad \quad
\end{center}
\caption{\it Zoom into the simulation results. (left) A cross section of the simulation box. The cells are distributed uniformly throughout the cell aggregate. The color corresponds to the leaf-levels in the Octree data structure. (right) A zoom into the simulation box, cells are colored by their transmembrane potential difference.} \label{fig::zoom}
\end{figure}

The different parameters defining the geometry and properties of the cells are tabulated in table \ref{tab:properties}. The computational configuration used to run this simulation is tabulated in table \ref{tab:config}. The resulting cell aggregate is illustrated in figure \ref{fig::clusterconfig}, with figure \ref{subfig::clusterPhi} depicting the electric potential (the aforementioned $u$ field) across the domain and figure \ref{subfig::clusterProcess} showing the partitioning between the $2048$ processors (identified with different colors - for visualization purposes, every adjacent 8 processors are displayed with same color). Figure \ref{fig::zoom} provides a cross section of the domain as well as a zoom that demonstrates that the cells are well-resolved. 

\begin{center}
\begin{tabular}{ |l|c|c|c| } 
\hline
Property & Symbol & Value & Units \\
\hline
\hline
Average cell radius &  $R_0$ & 7 & $\mu m$\\ 
 \hline
 \multicolumn{4}{| c |}{Cell geometric parameters range}\\
 \hline
 Cell radii & $r_0$ & 0.57-1.43 $\times R_0$ & $\mu m$\\
semi-axes & a, b, c & 0.8-1.2  $\times R_0$ & $\mu m$\\
\hline
 \multicolumn{4}{| c |}{Membrane electric parameters}\\
 \hline
Capacitance								& $C$  				& $9.5\times 10^{-3}$ & $F/m^2$ \\
Extracellular conductivity				& $\sigma^e$ 	& $15$ 						& $S/m$ \\
Intracellular conductivity 			& $\sigma^c$ 	& $1$ 						& $S/m$ \\
Voltage threshold for poration 				& $V_{ep}$ 	& $258\times 10^{-3}$		&$V$ \\
Membrane surface conductivity			& $S_0$ 	& $1.9$ 						& $S/m$ \\
Porated membrane conductance			& $S_1$   & $1.1\times10^6$			& $S/m^2$\\
Permeabilized membrane conductance			& $S_2$    & $10^4$			& $S/m^2$\\
Poration timescale									& $\tau_{ep}$    & $10^{-6}$		& $s$	\\
Permeabilization timescale									& $\tau_{perm}$    & $80\times10^{-6}$		& $s$	\\
Resealing timescale									& $\tau_{res}$    & $60$		& $s$	\\
Threshold for poration					& $X_{ep}$			& $0.5$		& -\\
\hline
 \multicolumn{4}{| c |}{Imposed electric pulse}\\
 \hline
 Electric field magnitude & $\vert \vect{E}\vert$	& $40$ & $kV/m$\\
 \hline
\end{tabular} 
\captionof{table}{Parameters of our simulation.}
\label{tab:properties}
\end{center}

\begin{center}
\begin{tabular}{ |l|c|c|c| } 
\hline
Property & Value \\
\hline
\hline
Macromesh in x,y \& z directions	$n_x\times  n_y\times  n_z$	& $2\times 2\times 2$  \\
Minimum/Maximum levels of refinement $(l_{min},l_{max})$		& $2\times9$\\
Total number of voronoi cells & $224,218,754$\\
Total number of nodes & $194,666,253$ \\
Number of processors 	& 2048 \\
Total time of simulation   &  $\approx 9$ hours \\
Number of timesteps 	 & 44 \\
Total physical time of the simulation 	 & $2.25\ (\mu s)$ \\
 \hline
\end{tabular} 
\captionof{table}{Computational aspects of our simulation.}
\label{tab:config}
\end{center}

\subsection{Convergence test and mesh independence}
\label{sec::convergence}

\begin{figure}[H]
\begin{center}
\includegraphics[width=\textwidth]{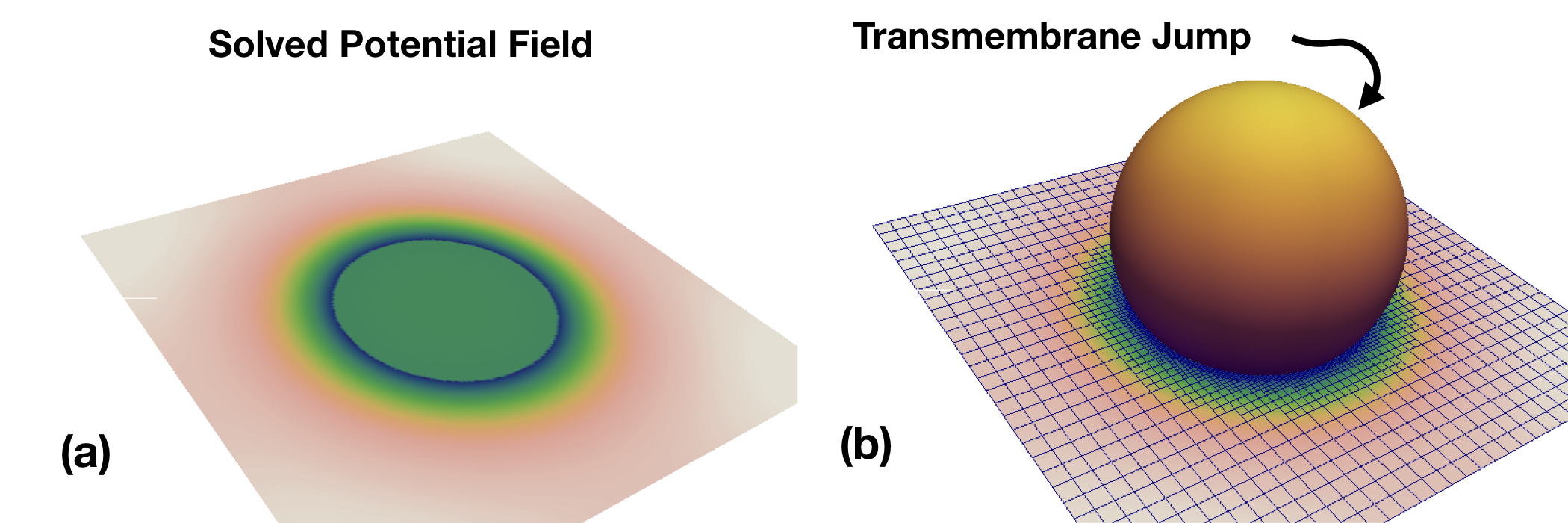} \label{subfig::JumpSol}
\end{center}
\caption{\it The configuration used for convergence tests. (a) A circular cross section of the cell demonstrates how the electric potential field experiences a jump when passing through the interface. (b) The jump is measured on the Octree mesh by first extrapolating solutions on each side to the opposite side and then subtracting the extrapolated values on the nodes around the interface.} \label{fig::Jump}
\end{figure}

To validate the numerical reliability of our implementation, we investigate the spatio-temporal convergence of the transmembrane potential jump, which is the key variable that couples the electropermeabilization equations. For this purpose, we consider a single spherical cell and track the evolution of the transmembrane potential jump $\left[u\right]$ at a $\pi/4$ radian distance from the cell's equator over time. Figure \ref{fig::Jump} illustrates the setup used for this purpose, as well as the refined mesh used.  We use the dynamic linear case with $S=S_L$, for which the transmembrane jump, $[u]$, satisfies:
\begin{equation}
C\frac{\partial [u]}{\partial t} + S_L[u]=\sigma_c\frac{\partial u}{\partial \vect{n}}.
\end{equation}
In this case, the exact solution is available for our validations and reads:
\begin{equation}
[u](t,\theta)=\frac{A}{S_L-B}g \bigg(1-e^{-\frac{S_L-B}{C}t}\bigg)\cos(\theta),
\end{equation}
where $g=ER_2$ and $\theta$ is the polar angle measured from the north pole. Also, $A$ and $B$ are given by:
\begin{subequations}
\begin{align} 
K^{-1}&=R_1^3(\sigma_e-\sigma_c)+R_2^3(2\sigma_e+\sigma_c),\\
A&=3\sigma_c\sigma_e R_2^2K,\\
B&=-\sigma_c\sigma_e(R_1^2+\frac{2R_2^3}{R_1}K).
\end{align} 
\end{subequations}
In our tests, we use $R_1=50\mu m$ and $R_2=600\mu m$.

\begin{figure}[H]
\begin{center}
\subfigure[]{\includegraphics[height=.45\textwidth]{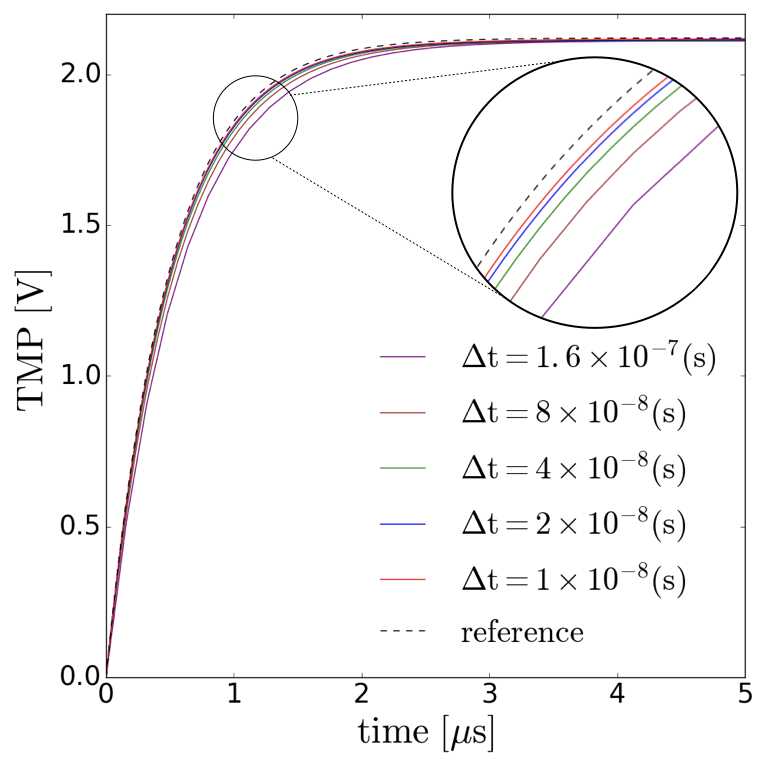} \label{subfig::convergence_time}} \quad \quad
\subfigure[]{\includegraphics[height=.45\textwidth]{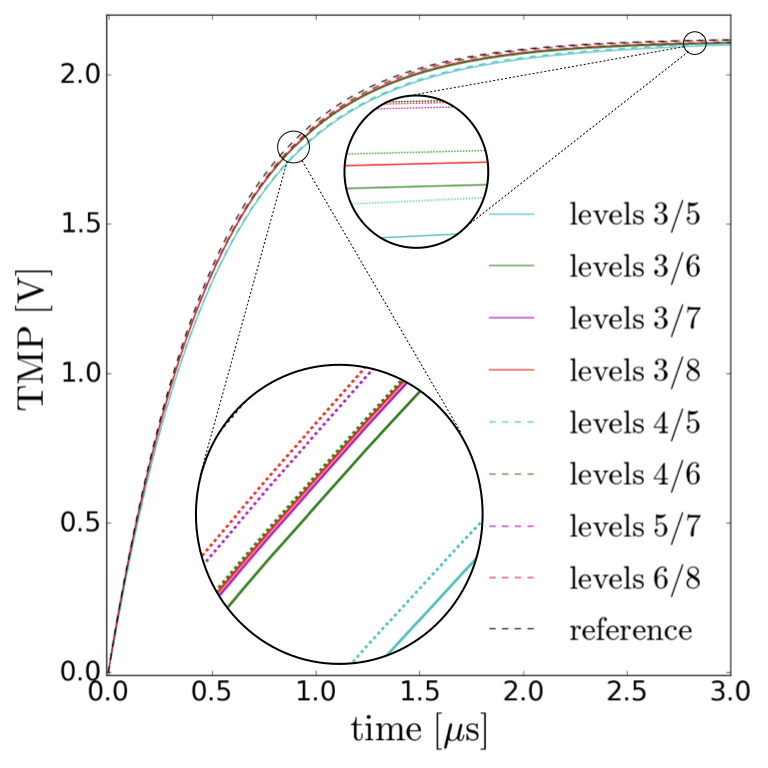} \label{subfig::convergence_space}} \quad \quad
\subfigure[]{\includegraphics[height=.45\textwidth]{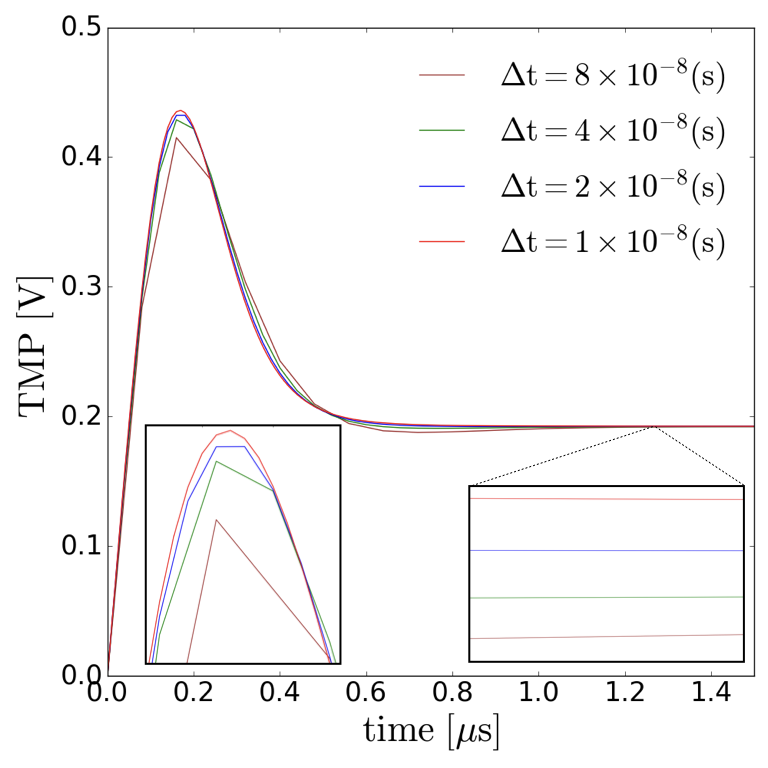} \label{subfig::convergence_time_nonlinear}} \quad \quad
\subfigure[]{\includegraphics[height=.45\textwidth]{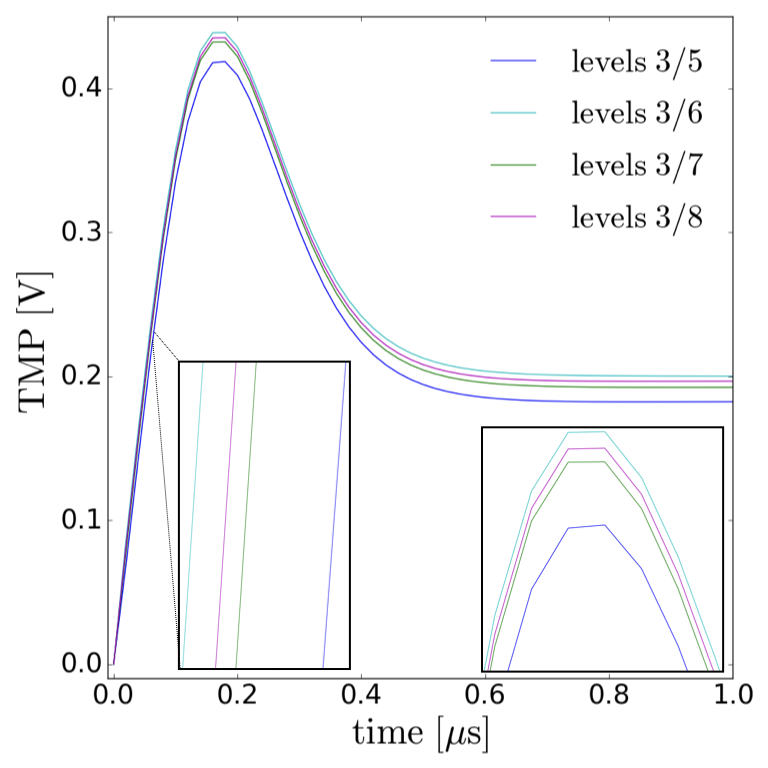} \label{subfig::convergence_space_nonlinear}} \quad \quad
\end{center}
\caption{\it Convergence analysis of section \ref{sec::convergence}. Figure (a) illustrates the temporal convergence of the TMP for five different time steps at a fixed grid size. Figure (b) demonstrate convergence in space consistent with the exact solution. Figures (c,d) are the temporal and spatial convergence for the full nonlinear case, respectively. Zoom-in figures are included in each plot for clarity.} \label{fig::independency}
\end{figure}

We perform the spatial and temporal refinements separately. First, we compare the results from simulations with different timesteps at a fixed resolution level of $\rm (l_{min}, l_{max})=(3, 7)$. In figure \ref{subfig::convergence_time} we show how the jump converges as we decrease the time step by a factor of $2$ each time. We performed our simulations with time steps of $\Delta t = 1\times 10^{-8}\ (s), 2\times 10^{-8}\ (s), 4\times10^{-8}, 8\times10^{-8}\ (s)$ and only for the linear case also with $1.6\times10^{-7}\ (s)$. This is because in the nonlinear case the latter time step is too big to capture the width of the peak in the jump profile. 
Also, in figure \ref{subfig::convergence_space} we increase the maximum refinement level while keeping the minimum refinement level fixed at $\rm l_{min}=3$ and the time step constant at $\Delta t = 2\times 10^{-8}\ (s)$; these are plotted with solid lines. Additionally, we perform identical simulations while simultaneously increasing both the minimum and maximum levels of refinements; these are shown with dashed lines. This is motivated by the observation that the solid lines in figure \ref{subfig::convergence_space}, corresponding to a fixed $\rm l_{min}=3$, converge to the exact solution at slower rate than the dashed lines. Maintaining low $\rm l_{min}$ while enhancing resolutions at the interface does not improve accuracy because errors produced at coarser grids far from interface become dominant in the simulation box, making further refinements useless when considering the error in the maximum norm. Even though both cases demonstrate convergence, increasing both the minimum and maximum refinement levels naturally exhibits a better convergence behavior. 

We also demonstrate that for the full nonlinear dynamic case, the convergence of our numerical results is achieved both in time and space in figures \ref{subfig::convergence_time_nonlinear} and \ref{subfig::convergence_space_nonlinear} respectively. In the nonlinear case, we choose a constant electric field intensity of $E=40kV/m$ across the domain in the $z$-direction. The size of the domain is $400\mu m$ in each spatial dimension. For the temporal convergence, we performed our simulations at fixed resolution levels of $(3,7)$ and for the spatial convergence we picked a fixed timestep of $\Delta t=2\times 10^{-8}(s)$ while varying the maximum refinement level. 

In the nonlinear case, convergence in time seems more problematic. As noted in \citep{guittet2017voronoi}, this is expected due to the highly nonlinear temporal nature of the equations, while the equations are spatially well-behaved. This implies that smaller timesteps are preferable over finer spatial resolutions for decreasing the numerical errors. Hence, we observe the system's response converges both in linear and nonlinear cases. We also note that in real case simulations that we perform the timestep is determined after setting the mesh at the desired resolution levels. Then in each simulation, the time-step is determined from $\Delta t = \Delta \vect{x}_{min}/dt_{scaling}$.

\subsection{Performance and scalability of the approach}
\label{sec::results_scaling}
\begin{figure}[H]
\begin{center}
\subfigure[]{\includegraphics[height=.45\textwidth]{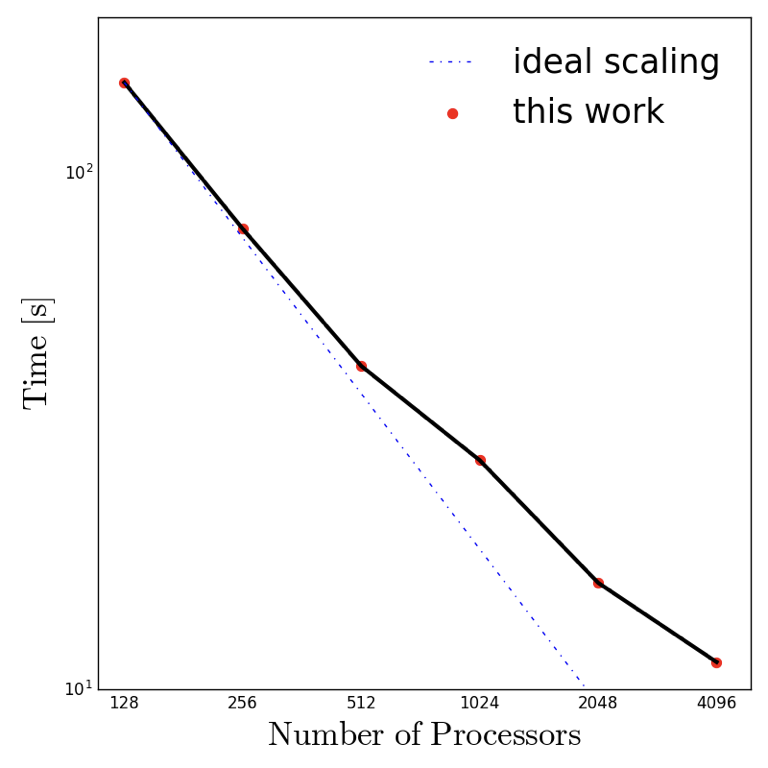} \label{subfig::clusterSm}} \quad \quad
\subfigure[]{\includegraphics[height=.45\textwidth]{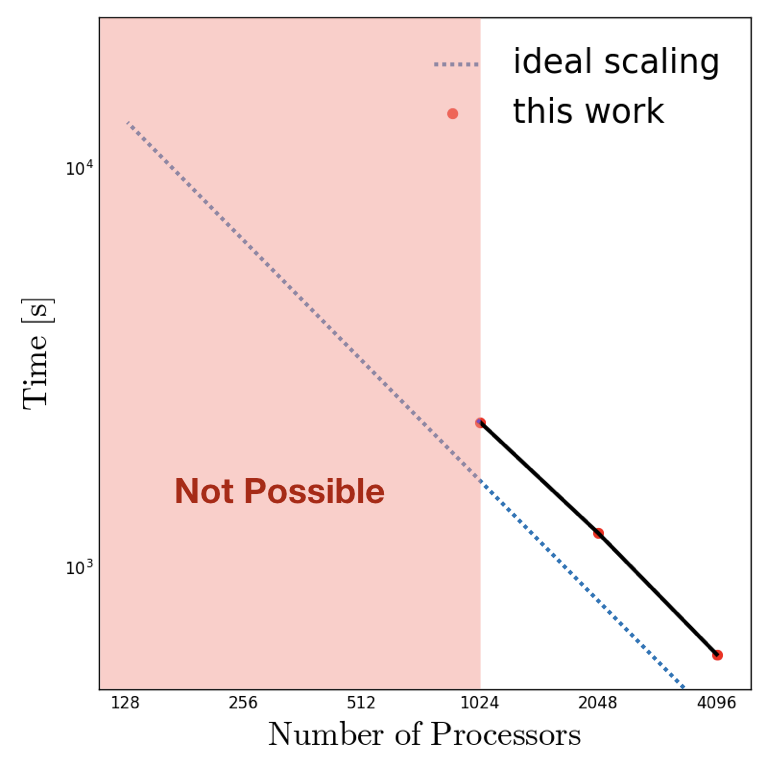} \label{subfig::TopPole}} \quad \quad
\end{center}
\caption{\it Scaling of the wall-clock time when increasing the number of processing cores. In both cases, the size of the problem is fixed and only the number of processors varies. The ideal scaling is shown with the dashed blue line. Our algorithms scale well for both small and large simulations. (left) A small simulation with $2,837,427$ nodes at levels (2,9) containing 313 biological cells. (right) Large aggregate with over $228,000,000$ nodes containing $31,320$ biological cells. In this case it is not possible to simulate large aggregates on small number of processors due to memory limitations, which we refer to as ``Not Possible''.} \label{fig::scaling}
\end{figure}

We show a simple test of the performance of the parallel approach for real applications of interest. We solve the same cell aggregate problem introduced in section \ref{sec::NumericalResults} on different numbers of processors while keeping all other parameters fixed. This test captures the full problem complexity and hence enables a reasonable assessment of the computational efficiency and scalability of the approach. Constructing the Voronoi mesh at each time step and solving the linear system arising from the discretization introduced in section \ref{subsec:discretize} constitute the bulk of the computational expense of our approach. Figure \ref{fig::scaling} demonstrates that our approach tackles these tasks excellently up to $4096$ processors, which is the upper limit to our current account on the ``Stampede2'' supercomputer. 

In figure \ref{fig::scaling}, we also show the scaling test for a smaller cell density in order to demonstrate the capabilities of our implementation at lower problem sizes, where communication overhead easily exceeds that of computational time. Interestingly, we find that our approach exhibits excellent scalability even for quite small problems.

We should emphasize that parallelization is only one avenue to simulating larger problems in our methodology. Another significant aspect is the use of adaptive mesh refinement using Octree grids. This introduces a significant reduction in the size of the grid from $\approx2^{30}$ nodes to $194,666,253$ nodes in this example. We refer the interested reader to \cite{Mistani2017} for a quantitative study of this enhancement. This consequently advances the limits of the possible simulation scales with the current state-of-the-art available resources. 

\section{Mesoscale Phenomenology}
\label{sec::emergent}
Cell aggregates are complex systems composed of many cells that each follow a set of principles and collectively reach an equilibrium state with their environment. Cell aggregates exhibit emergent phenomena \cite{nagel1961structure}, \textit{i.e.} ``novel and robust behaviors of a system that appear at the limit of some parameter in the system'' \cite{butterfield2011less,butterfield2011emergence}. In our case, a weak form of emergence appears at some finite limit of system size. These novel features are robust against certain details at the smaller scales of the aggregate; viz. in the sense that via the process of coarse-graining the renormalized parameters describing theories at different scales \textit{always} converge to certain fixed values in natural systems (cf. \cite{kadanoff2013relating}). This generic feature of complex systems is recognized as a fundamental principle of nature \cite{butterfield2015renormalization}. Recently the descriptive framework that arises by relying on this aspect of complex systems has been discussed by \cite{Mistani2019a,Mistani2019b}.

In the study of complex systems, computational strategies provide powerful or in some specific cases the \textit{only} method to exploit the so called ``weak emergent'' phenomena, first described by Bedau 2002 \citep{bedau2002downward}. Weak emergence is attributed to those physical aspects of complex systems that, in practice, only appear through computer simulations. This is due to the nonlinearity of the micro-level equations and the complex interactions between its constituent parts. For a comprehensive review of this topic we refer the interested reader to Fulmer \etal \citep{fulmer2016convergence}.

As in most large-scale numerical simulations, our main purpose is to study the non-local effects that are not already encoded \textit{locally} in the governing partial differential equations, but are encrypted in the spatial domain as a whole and influence the overall behavior via feedback processes among elements. In the case of electroporation, such influences are in part due to the heterogeneous cell topologies, long range electrostatic interactions, and the overall shape of the aggregate among other factors. In this section, we aim to show that macro-level features of cell aggregates are recovered in our methodology. We first demonstrate the influence of cell shape on the macro-level properties of the aggregate, and will present first results for a tumor-like aggregate.

\subsection{Effect of biological cell shape}
\label{sec::shape}
\begin{figure}[H]
\begin{center}
\subfigure[]{\includegraphics[height=.33\textwidth]{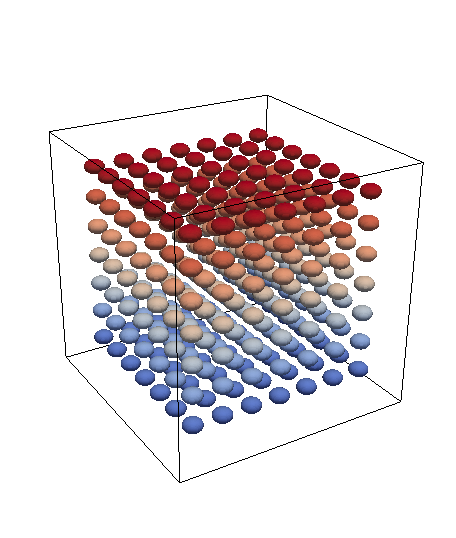} \label{subfig::oblate}} \quad \quad
\subfigure[]{\includegraphics[height=.33\textwidth]{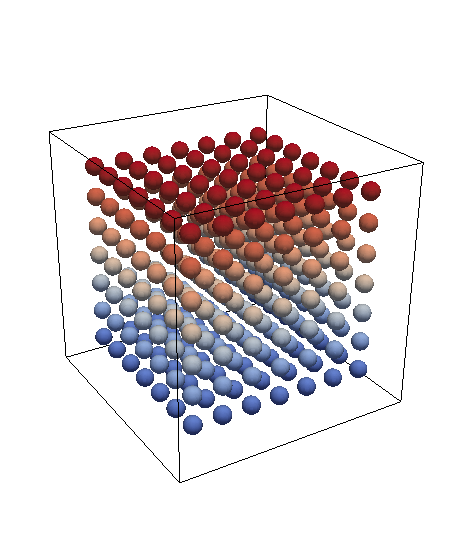} \label{subfig::sphere}} \quad \quad
\subfigure[]{\includegraphics[height=.33\textwidth]{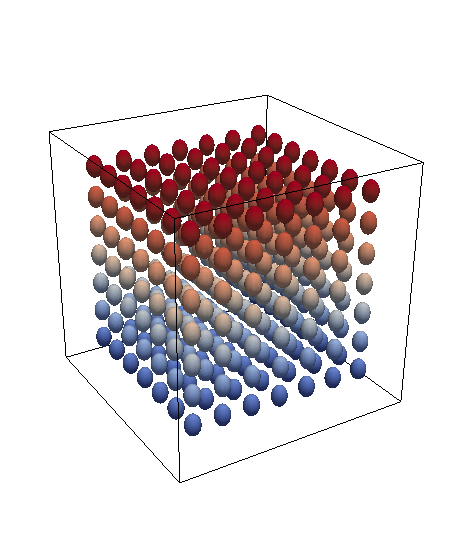} \label{subfig::prolate}} \quad \quad
\end{center}
\caption{\it Arrays of cells used in section \ref{sec::shape}. (left) oblates, (middle) spheres, and (right) prolates with equal volumes.} \label{fig::topog}
\end{figure}

\begin{figure}[H]
\begin{center}
\includegraphics[width=\textwidth]{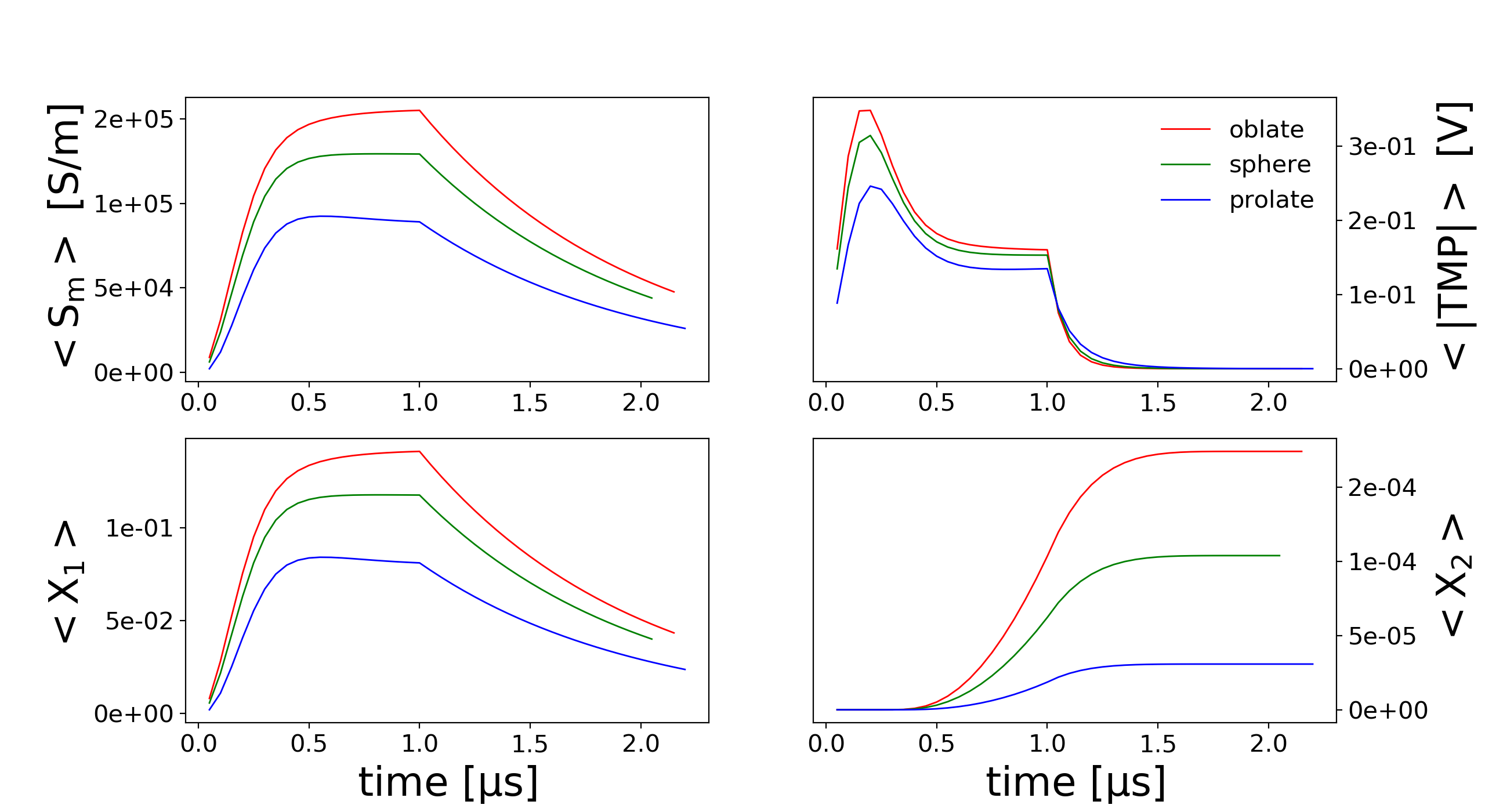} 
\end{center}
\caption{\it Effect of cell shape on the parameters of the electroporation model.} \label{fig::shapes}
\end{figure}

Biological cells come in different shapes. We place three simple types of cells in the same experimental setup and compare their bioelectric behavior. To this end, we choose to place oblate, spherical and prolate cells with identical volume on a $7\times7\times7$ regular lattice. Figure \ref{fig::topog} shows the configurations used in our experiments, and the effect of cell shape is compared in figure \ref{fig::shapes}. One can observe that cells with prolate topology exhibit higher levels of permeabilization, spheres fall in between and oblate spheroids are the least electroporated. This is consistent with previous reports of \cite{guittet2017voronoi}, and may be due to higher effective cross section area exposed to the influx of the electric field. 

\subsection{Shadowing effect}
\label{sec::shadowing}
\begin{figure}[H]
\begin{center}
\subfigure[]{\includegraphics[width=0.45\textwidth]{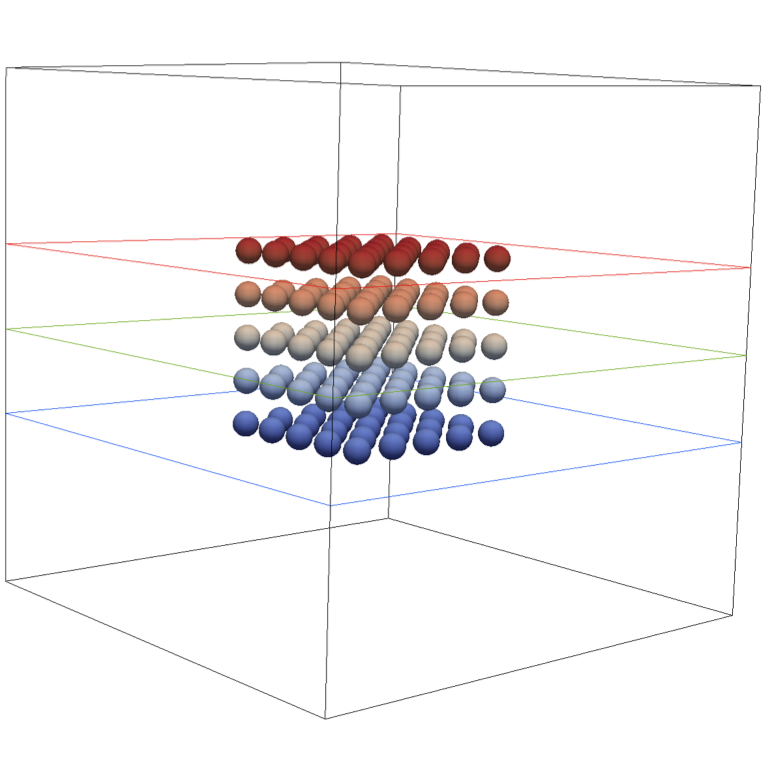} \label{subfig::shadows1}}\quad \quad
\subfigure[]{\includegraphics[width=0.45\textwidth]{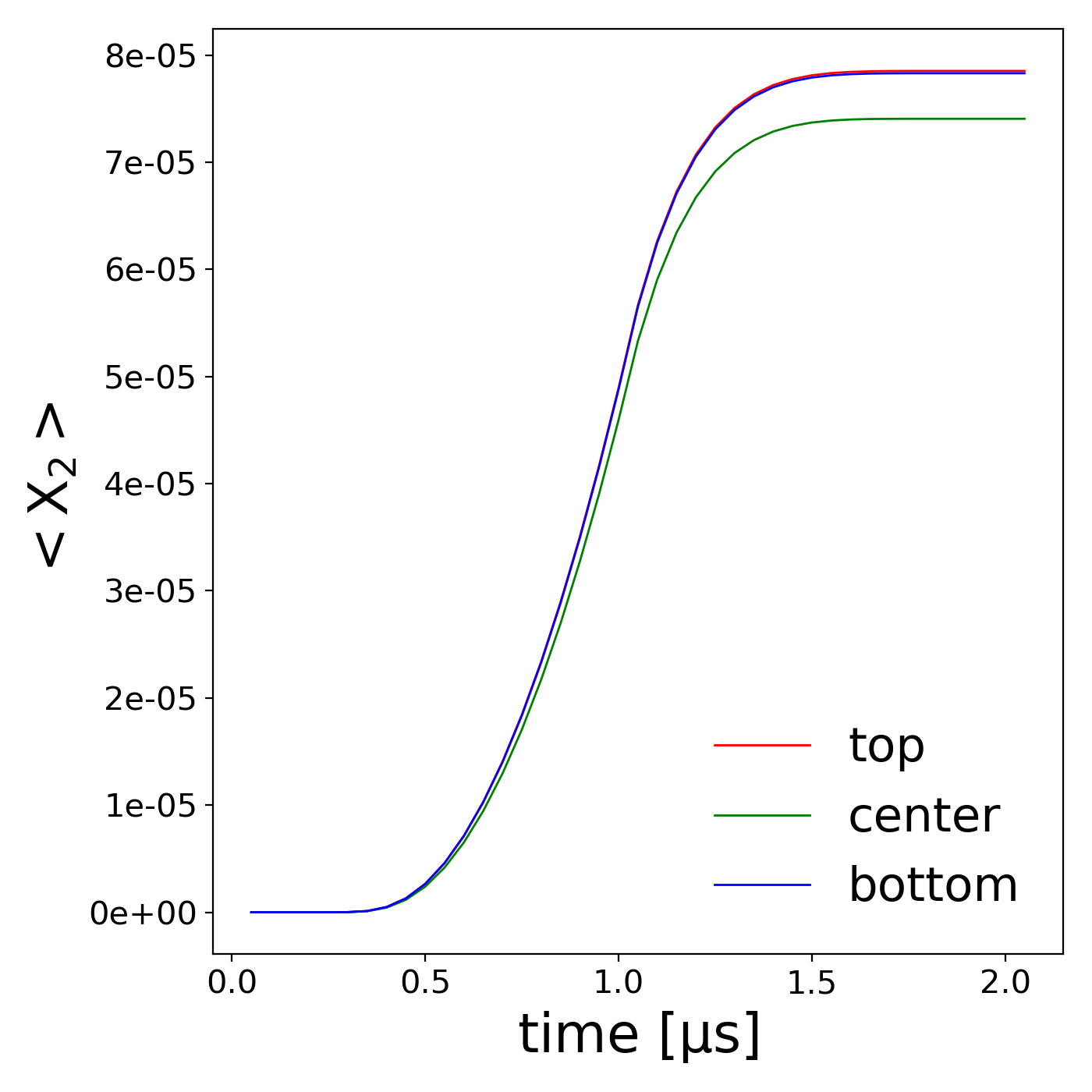} \label{subfig::shadows2}}\quad \quad
\end{center}
\caption{\it Shadowing effect.} \label{fig::shadows}
\end{figure}

Shadowing refers to the adverse effect of upstream cells to the permeabilization levels exhibited by their downstream counterparts. We performed experiments on a controlled sample of 125 spherical cells in a cubic lattice centered in a bounding box with twice the size of the lattice. We place cells symmetrically in a $5 \times 5 \times 5$ array as depicted in figure \ref{subfig::shadows1}. We compare the surface average of $X_2$ parameter over the surface of all cells in the top, center, and bottom rows. The results are given in figure  \ref{subfig::shadows2}. 

As expected the middle row is less permeabilized, and cells closer to the electrodes (in this specific configuration) exhibit higher levels of permeabilization. In particular, this observation is in accordance with the experimental data of spheroid electroporation of Rols \etal~\cite{WASUNGU2009278}. Note that owing to the reflection symmetry, top and bottom slices are in identical environments, this is also reflected by the overlapping measurements for their permeabilization curves as in figure \ref{subfig::shadows2}. 

So far we have only considered regular lattice configurations, in the remaining of this work, we focus on the tumor-like demonstration case that is depicted in figure \ref{fig::clusterconfig}. To date, studying computationally this relevant biological structure is only possible with the computational approach introduced in this manuscript.

\subsection{Electroporation fraction}
\label{sec::fraction}
In experiments, one can measure the fraction of cells that are electropermeabilized more than a detectable threshold. In order to compare our numerical results with experiments, we set the minimum detectable threshold for electropermeabilization to different values: $$\rm S_m\ge (100 \ or \ 1,000\ or \ 10,000\ or \ 100,000) S_L.$$ Then, we measure the fraction of total electropermeabilized surface area of all cells normalized by the total surface area of the cells. 

\begin{figure}[H]
\begin{center}
\subfigure[]{\includegraphics[width=.25\textwidth]{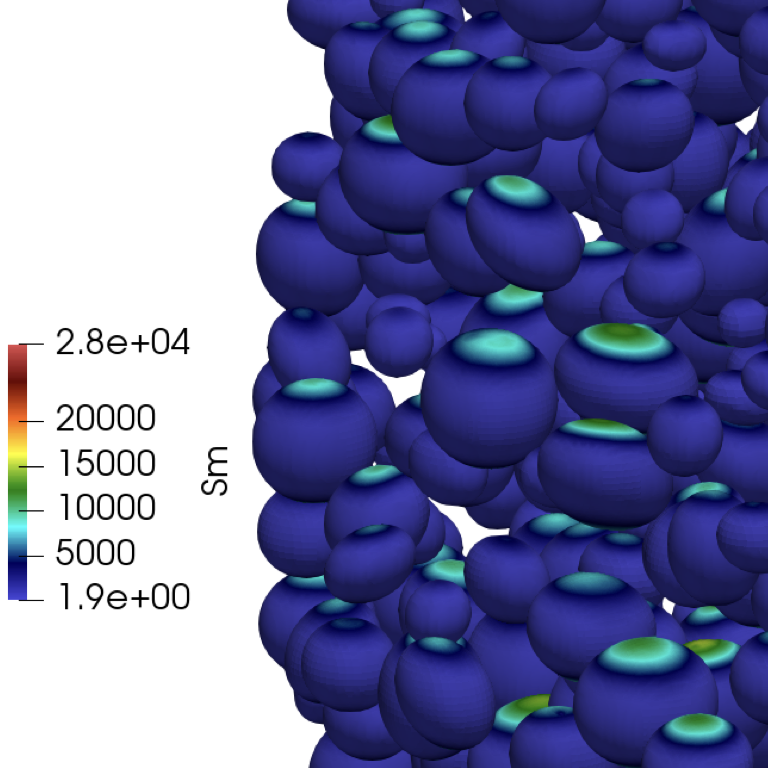} \label{subfig::TopPole}} \quad \quad
\subfigure[]{\includegraphics[width=.25\textwidth]{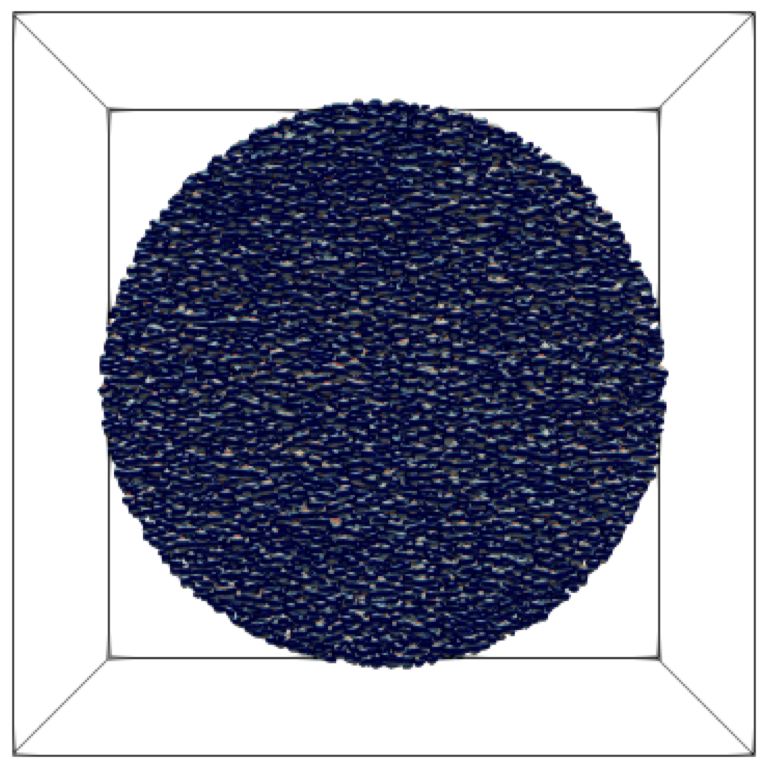} \label{subfig::TopPole}} \quad \quad
\subfigure[]{\includegraphics[width=.25\textwidth]{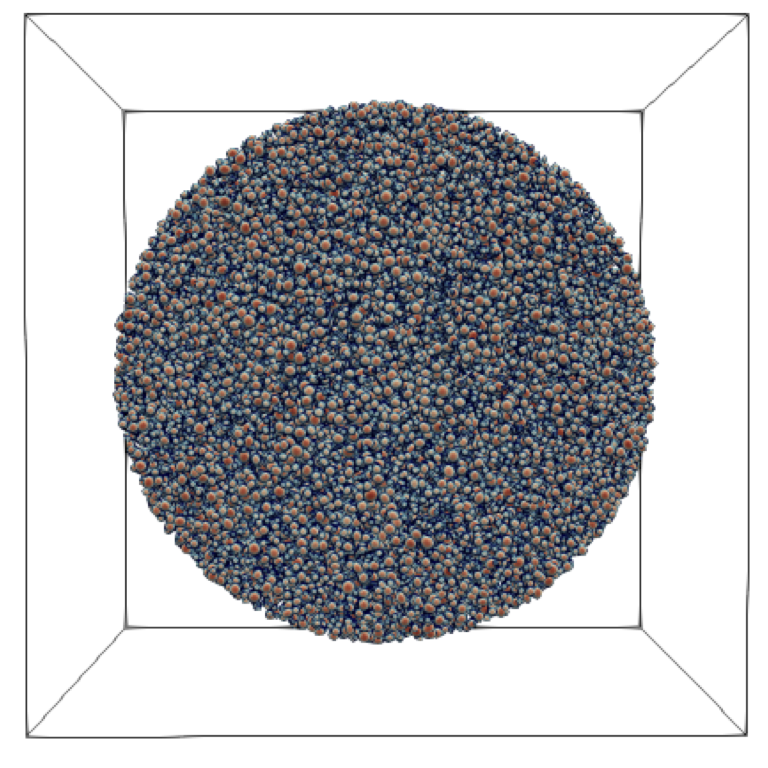} \label{subfig::TopPole}} \quad \quad
\end{center}
\caption{\it Permeabilization pattern in a heterogenous aggregate. (a) Cells are mostly permeabilized along their vertical ($z$-) axis, (b) is a side view of the aggregate, and (c) is a top-view of the cell aggregate. Hotter colors represent higher values of cell membrane conductance.} \label{fig::clusterPerm1}
\end{figure}

\begin{figure}[H]
\begin{center}
\includegraphics[width=\textwidth]{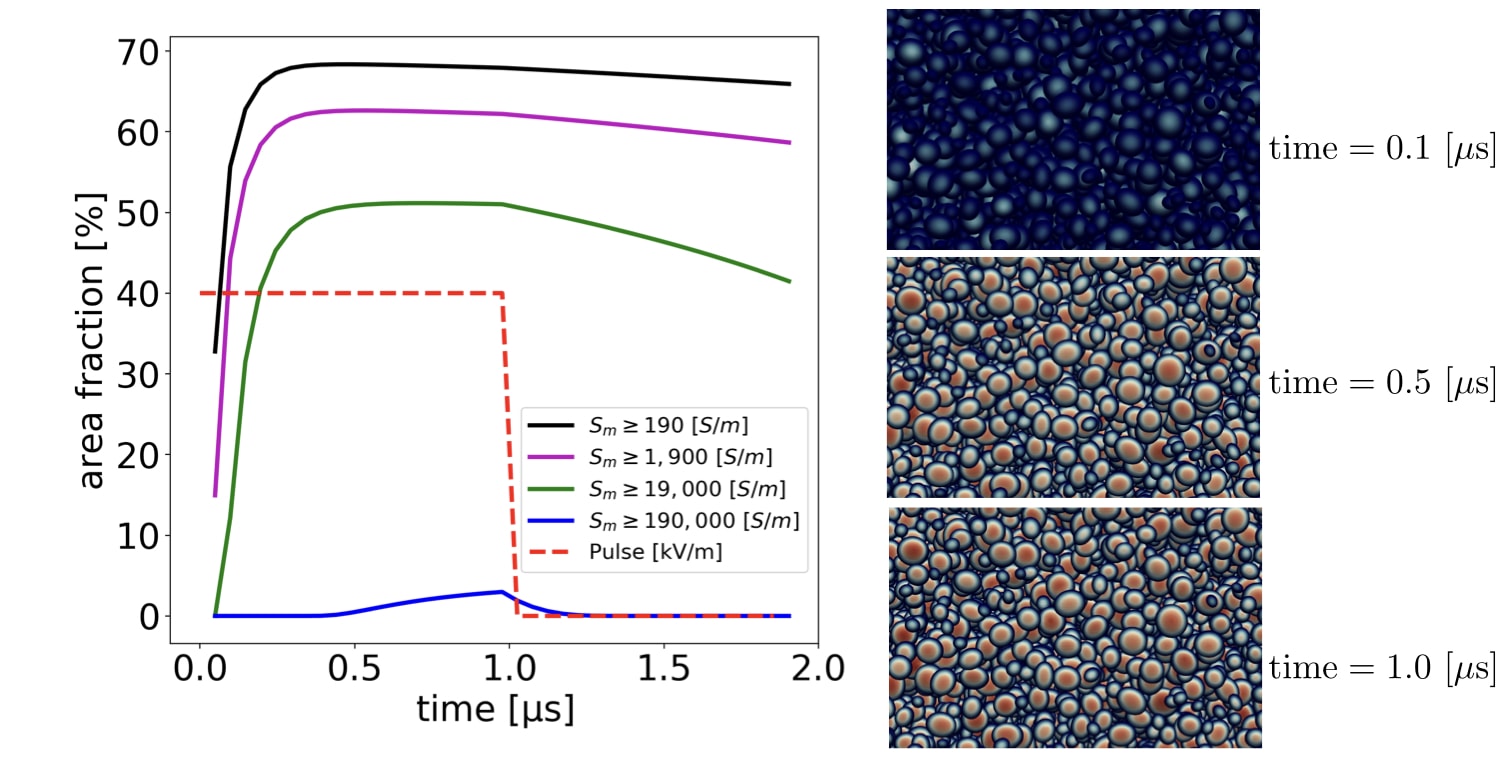} 
\end{center}
\caption{\it Electropermeabilization fraction over time for a $1\mu s$ square pulse of $40kV/m$. Figures on the right panel are color coded by conductance, with hotter colors encoding higher conductance levels.} \label{fig::area}
\end{figure}

Figure \ref{fig::clusterPerm1} depicts the permeabilization pattern throughout a dense suspension (volume fraction of $13\%$), and figure \ref{fig::area} quantifies the evolution of the membrane electropermeabilization fraction. Remarkably, we observe that the maximum value of this fraction under a short $40kV/m$ electric pulse reaches $\approx 70\%$, $ 65\%$, $ 50\%$ and $ 5\%$ for the given thresholds respectively. This is in qualitative agreement with the experimental results of Pucihar \etal \cite{pucihar2007electropermeabilization}.

\begin{figure}[H]
\begin{center}
\includegraphics[width=\textwidth]{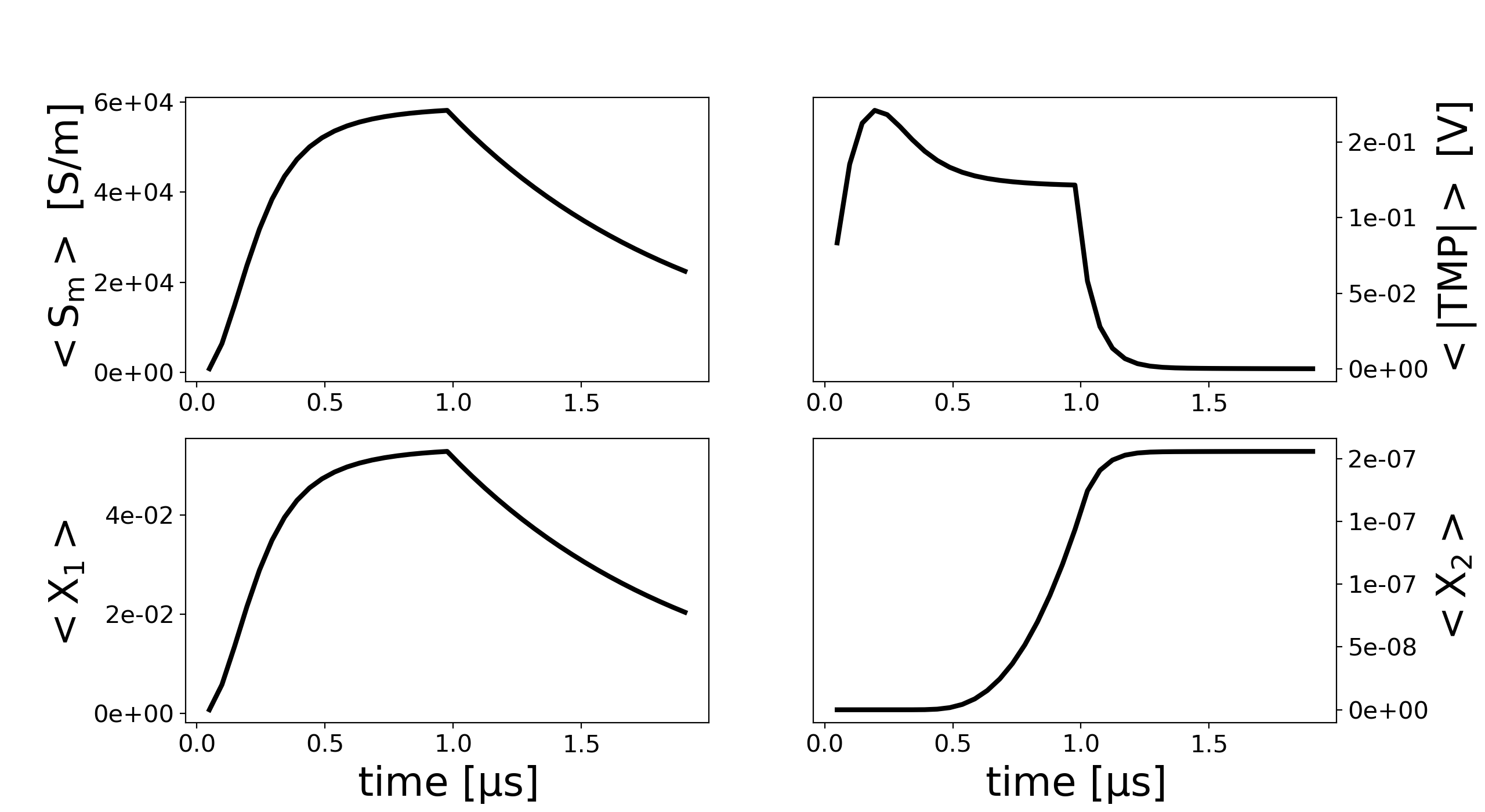} \label{subfig::params} \quad \quad
\end{center}
\caption{\it Time evolution of relevant parameters averaged over the membranes of the $27,440$ cells in our simulations. The applied pulse is turned off at $1\mu s$.} \label{fig::allparams}
\end{figure}

The evolution of the relevant electropermeabilization parameters including membrane conductance ($S_m$), level of membrane poration ($X_1$), level of membrane permeabilization ($X_2$) and absolute value of the transmembrane potential (TMP) are shown in figure \ref{fig::allparams} for reference. One observation is that the transmembrane voltage does not vanish spontaneously after the external pulse is turned off; this is due to the capacitive nature of the cell membranes that maintain a slowly vanishing electric field in the environment.

\begin{figure}[H]
\begin{center}
\subfigure{\includegraphics[width=\textwidth]{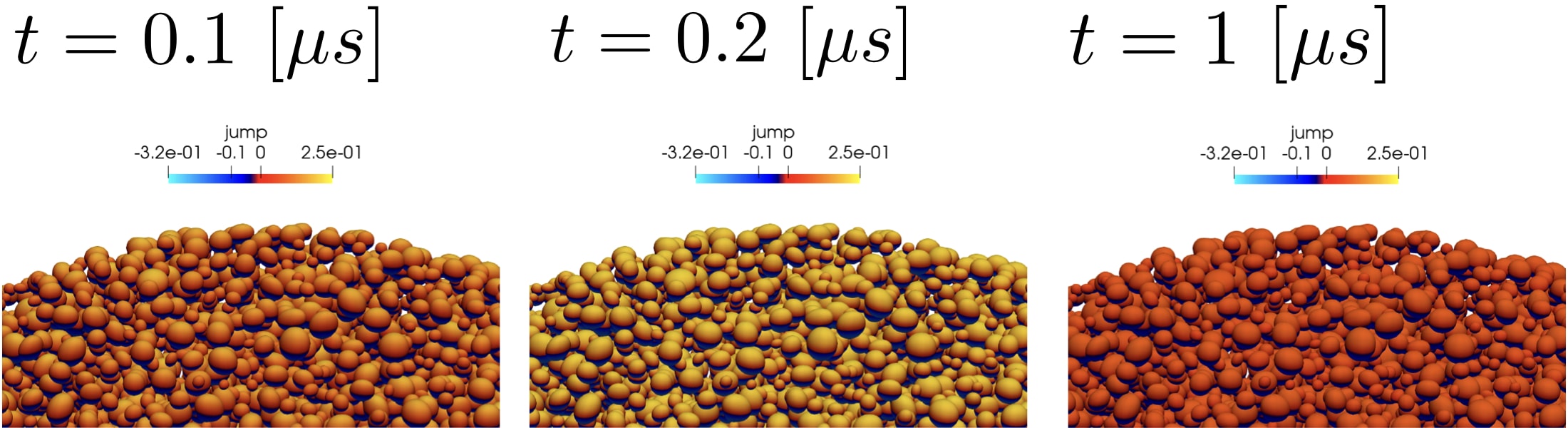} \label{subfig::VnEvolution}} \quad \quad
\end{center}
\caption{\it Time evolution of the transmembrane potential in a cell aggregate. Hotter colors correspond to higher values of transmembrane voltage.} \label{fig::VnEvolution}
\end{figure}

The signature of the nonlinear model underlying the evolution of the transmembrane voltage is also evident in these figures. We present three snapshots of the transmembrane potential in the aggregate in figure \ref{fig::VnEvolution}. These snapshots capture the initial overshoot in the transmembrane voltage (cf. figure \ref{fig::allparams}) and then the saturation phase that follows. These snapshots are color coded according to the transmembrane potential. 

\subsection{Impedance of the aggregate}
\label{sec::impedance}
\begin{figure}[H]
\begin{center}
\subfigure{\includegraphics[height=.45\textwidth]{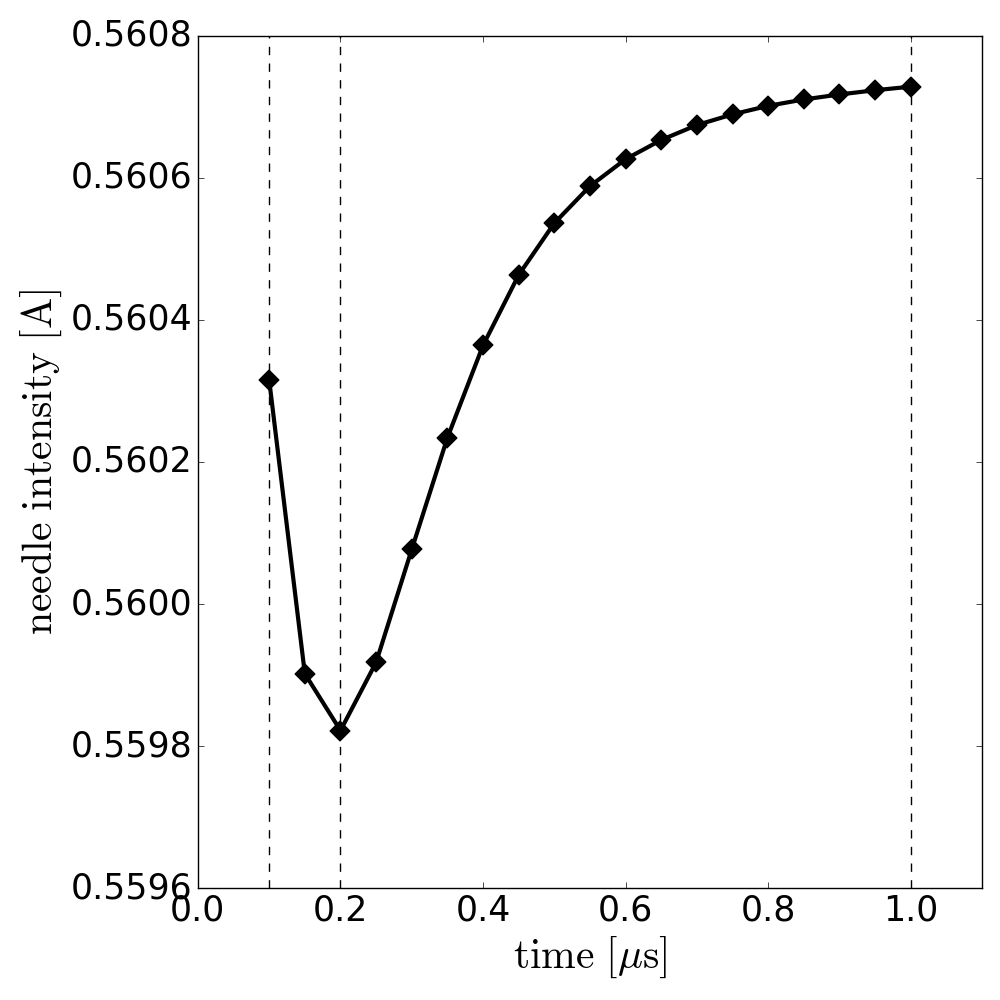} \label{subfig::intensity}} \quad \quad
\subfigure{\includegraphics[height=.45\textwidth]{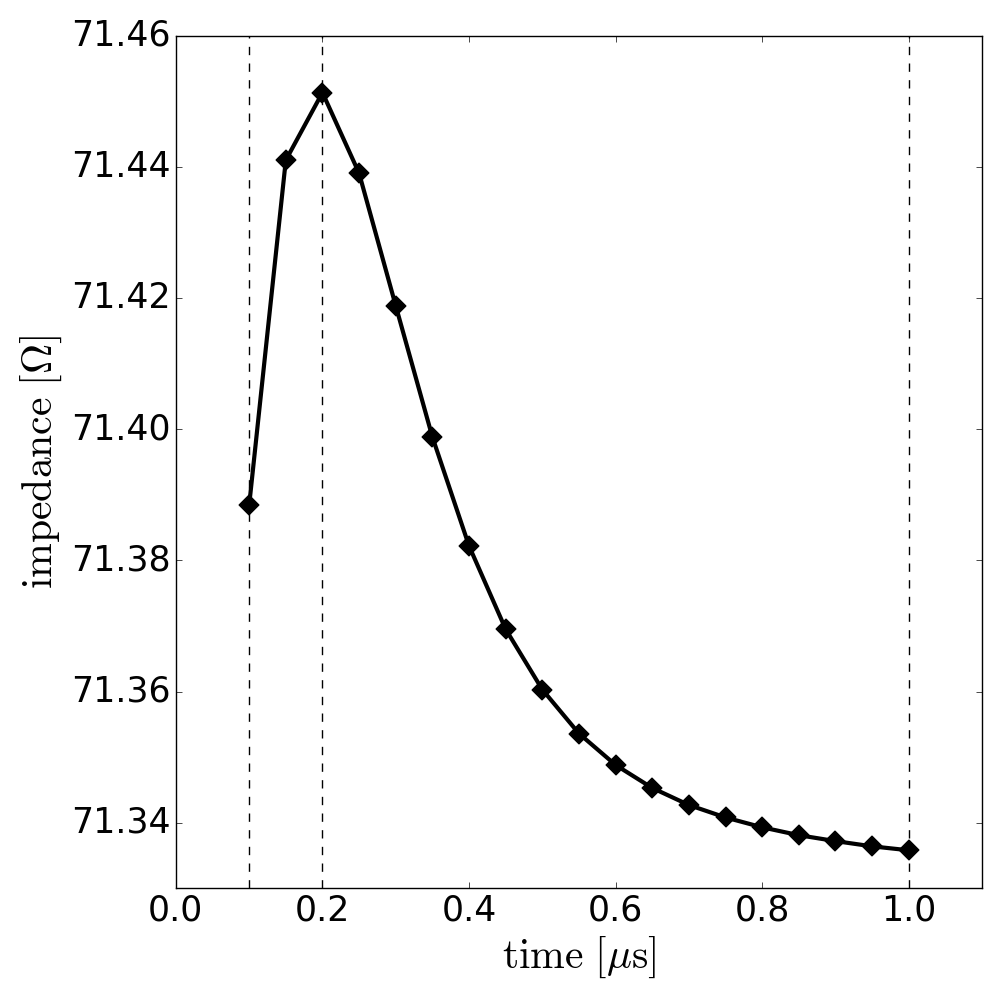} \label{subfig::impedance}} \quad \quad
\end{center}
\caption{\it (left) time evolution of the needle intensity, as well as (right) the resulting aggregate impedance under a constant external potential difference. Vertical dashed lines mark the times $t=0.1\ [\mu s]$, $t=0.2\ [\mu s]$ and $t=1\ [\mu s]$ for which the corresponding transmembrane voltages are visualized in figure \ref{fig::VnEvolution}.} \label{fig::impedance}
\end{figure}

In these simulations we apply a constant and uniform potential difference between the electrodes. The electric field will adapt to the geometrical configuration of the domain as well as the cells, while the cell membranes also distort the field. The distortions in the observed electric field close to the boundaries, where the electrodes are located, produce a different profile for the ``needle potential'' that the cell aggregate experiences. Needle intensity is defined as:
 \begin{equation}
I(t) = \int_{\mathcal{E}_1} \sigma^e \partial_{\vect{n}} V(t,x)\cdot \vect{n} ds,
\label{eq::intensity}
\end{equation}
where $\mathcal{E}$ is one of the electrodes where the voltage is imposed. The evolution of the needle intensity for the tumor-like aggregate is shown in figure \ref{subfig::intensity}.

Furthermore, one can measure the overall permeability within the environment by measuring the impedance of the sample detected at the electrodes. We define the impedance of the cell aggregate as:
\begin{equation}
Z(t) = \frac{\int_{\mathcal{E}_{1-2}}V(t,x) ds/\int_{\mathcal{E}_{1}}ds}{ \int_{\mathcal{E}_1}\sigma \partial_{\vect{n}} V(t,x)\cdot \vect{n} \, ds},
\label{eq::impedance}
\end{equation}
where $\mathcal{E}_{1}$ and $\mathcal{E}_{2}$ are either the top or the bottom electrode, and $\mathcal{E}_{1-2}$ is the difference of the integral between $\mathcal{E}_{1}$ and $\mathcal{E}_{2}$ electrodes. Note that the exact choice of labels does not change the result due to continuity of current through the medium.

The time evolution of the impedance of the aggregate is shown in figure \ref{subfig::impedance}. Comparison with figure \ref{fig::allparams} suggests a strong negative correlation between impedance and the overall degree of permeability. We find that even though permeabilized cells have a huge increase of their membrane conductance (from $1$ to $10^4$ $S/m^2$), as illustrated in figure \ref{fig::allparams}, the relative impedance of the aggregate drops about $\approx 0.15\%$ after $1\mu s$ of a constant external electric pulse.

\section{Conclusion} \label{sec::Conclusion}

We have presented a computational framework for parallel simulations of cell aggregate electropermeabilization at the mesoscale. We used an adaptive Octree/Voronoi mesh along with a numerical treatment that preserves the jump in the electric potential across each cell's membrane. The core aspects of our methodology are its efficiency and excellent scalability, making it possible to consider meaningful simulations of tumor-like spheroids, as opposed to previous serial approaches that were not able to go beyond micro-scale simulations. We have presented preliminary numerical results on cell aggregate electropermeabilization that are in qualitative agreement with experimental observations. This work thus paves the way for a wide range of comparisons with biological experiments, as it makes possible the multiscale understanding of electroporation from the cell to the tissue.


\section*{Acknowledgement}
The research of P. Mistani, A. Guittet and F. Gibou was supported by NSF DMS-1620471 and ARO W911NF-16-1-0136.  C. Poignard research is supported by Plan Cancer DYNAMO (ref. PC201515) and Plan Cancer NUMEP (ref. PC201615). P. Mistani would like to thank Daniil Bochkov in the CASL group for fruitful discussions that have contributed to this research. This work used the Extreme Science and Engineering Discovery Environment (XSEDE), which is supported by National Science Foundation grant number ACI-1053575. The authors acknowledge the Texas Advanced Computing Center (TACC) at The University of Texas at Austin for providing HPC and visualization resources that have contributed to the research results reported within this paper.  This research was performed in part within the scope of the Inria associate team NUM4SEP, between the CASL group at UCSB and the Inria team MONC. C.P.'s research is partly performed within the scope of the European Associated Laboratory EBAM on electroporation, granted by CNRS.
\section*{References}
\bibliographystyle{abbrv}
\addcontentsline{toc}{section}{\refname}
\bibliography{references_electroporation}

\end{document}